\documentclass[11pt]{article}
\usepackage{a4wide}
\usepackage{amsmath,amssymb,amsfonts}
\usepackage{comment}
\usepackage{epsfig,verbatim,graphics}
\usepackage{subfigure,slashed}
\usepackage{hyperref}

\newcommand{\mathsym}[1]{{}}

\newcommand{\eref}[1]{(\ref{#1})}

\renewcommand\({\left(}
\renewcommand\){\right)}
\renewcommand\[{\left[}
\renewcommand\]{\right]}

\newcommand{\dd}{{\rm d}}
\declareslashed{}{{^-}}{.1}{0.1}{\rm d}

\newcommand{\e}{{\rm e}}

\newcommand\vp{\varphi}
\newcommand\eps{\epsilon}
\newcommand\mpl{m_{\rm p}}

\newcommand\mn{{\mu\nu}}
\newcommand\xg{{\xi_G}}

\def\ghh{\gamma^{hh}}

\def\ba{\begin{eqnarray}}
\def\ea{\end{eqnarray}}
\def\be{\begin{equation}}
\def\ee{\end{equation}}

\def\fac{s}

\def\L{\mathcal{L}}
\def\O{\mathcal{O}}

\def\nn{\nonumber}
\def\({\left(}
\def\){\right)}

\def\eref#1{(\ref{#1})}

\usepackage{tikz}
\usetikzlibrary{arrows,shapes}
\usetikzlibrary{trees}
\usetikzlibrary{matrix,arrows} 				% For commutative diagram
\usetikzlibrary{positioning}				% For "above of=" commands
\usetikzlibrary{calc,through}				% For coordinates
\usetikzlibrary{decorations.pathreplacing}  % For curly braces
\usepackage{pgffor}							% For repeating patterns

\usetikzlibrary{decorations.pathmorphing}	% For Feynman Diagrams
\usetikzlibrary{decorations.markings}
\tikzset{
    vector/.style={decorate, decoration={snake}, draw},
	provector/.style={decorate, decoration={snake,amplitude=2.5pt}, draw},
	antivector/.style={decorate, decoration={snake,amplitude=-2.5pt}, draw},
    fermion/.style={draw=black, postaction={decorate},
        decoration={markings,mark=at position .55 with {\arrow[draw=black]{>}}}},
    fermionbar/.style={draw=black, postaction={decorate},
        decoration={markings,mark=at position .55 with {\arrow[draw=black]{<}}}},
    fermionnoarrow/.style={draw=black},
    gluon/.style={decorate, draw=black,
        decoration={coil,amplitude=4pt, segment length=5pt}},
    scalar/.style={dashed,draw=black, postaction={decorate},
        decoration={markings,mark=at position .55 with {\arrow[draw=black]{>}}}},
    scalarbar/.style={dashed,draw=black, postaction={decorate},
        decoration={markings,mark=at position .55 with {\arrow[draw=black]{<}}}},
    scalarnoarrow/.style={dashed,draw=black},
    electron/.style={draw=black, postaction={decorate},
        decoration={markings,mark=at position .55 with {\arrow[draw=black]{>}}}},
	bigvector/.style={decorate, decoration={snake,amplitude=4pt}, draw},
}

\newcommand{\roughly}[1]{\mathrel{\raise.3ex\hbox{$#1$\kern-0.85em
\lower1ex\hbox{$\sim$}}}}

\begin{document}

\begin{titlepage}%1
\begin{center}

\today
\hfill NIKHEF 2015-026 \\
\hfill DAMTP-2015-45

\vskip 1.5cm

{\LARGE \bf Quantum corrections in Higgs inflation: the Standard Model case}

\vskip 1cm

% this command makes the email footnote's into symbols
\renewcommand*{\thefootnote}{\fnsymbol{footnote}}
\setcounter{footnote}{0}

{\bf
    Damien P.\ George$^{1,2}$\footnote{{\tt dpg39@cam.ac.uk},
        \footnotemark[2]{\tt sander.mooij@ing.uchile.cl},
        \footnotemark[3]{\tt mpostma@nikhef.nl}},
    Sander Mooij$^{3}$\footnotemark[2] and
    Marieke Postma$^4$\footnotemark[3]
}

% this command reverts the footnote style to back to number
\renewcommand*{\thefootnote}{\number{footnote}}
\setcounter{footnote}{0}

\vskip 25pt

{\em $^1$ \hskip -.1truecm
Department of Applied Mathematics and Theoretical Physics,\\
Centre for Mathematical Sciences, University of Cambridge,\\
Wilberforce Road, Cambridge CB3 0WA, United Kingdom
}

\vskip 20pt

{\em $^2$ \hskip -.1truecm
  Cavendish Laboratory, University of Cambridge,\\
  JJ Thomson Avenue, Cambridge CB3 0HE, United Kingdom }

\vskip 20pt

{\em $^3$ \hskip -.1truecm
FCFM, Universidad de Chile, \\Av. Blanco Encalada 2008, \\837.0415 Santiago, Chile
}

\vskip 20pt

{\em $^4$ \hskip -.1truecm
Nikhef, \\Science Park 105, \\1098 XG Amsterdam, The Netherlands
}

\end{center}

\vskip 0.5cm

\begin{center} {\bf ABSTRACT}\\[3ex] \end{center}
We compute the one-loop renormalization group equations for Standard Model Higgs inflation. The calculation is done in the Einstein frame, using a covariant formalism for the multi-field system. All counterterms, and thus the betafunctions, can be extracted from the radiative corrections to the two-point functions; the calculation of higher n-point functions then serves as a consistency check of the approach.  We find that the theory is renormalizable in the effective field theory sense in the small, mid and large field regime. In the large field regime our results differ slightly from those found in the literature, due to a different treatment of the Goldstone bosons. 
%Also, our approach to find the running of the non-minimal coupling is new.

\end{titlepage}

\newpage
\setcounter{page}{1} \tableofcontents

\newpage

%%%%%%%%%%%%%%%%%%%%%%%%%%%%%%%%%%%%%%%%%%%%%%%%%%%%%%%%%%%%%%
%%%%%%%%%%%%%%%%%%%%%%%%%%%%%%%%%%%%%%%%%%%%%%%%%%%%%%%%%%%%%%
%%%%%%%%%%%%%%%%%%%%%%%%%%%%%%%%%%%%%%%%%%%%%%%%%%%%%%%%%%%%%%

\section{Introduction}

In Higgs inflation the Higgs field of the Standard Model (SM) is
coupled non-minimally to gravity~\cite{Salopek:1988qh, bezrukov1,
  bezrukov2, bezrukov_loop, bezrukov3, bezrukov4}.  Apart from this
single non-minimal coupling, no new physics is needed to describe
inflation and the subsequent period of reheating, and the theory seems
to be extremely predictive. However, this presupposes the parameters
of the theory at high and low scale are related by renormalization
group (RG) flow. The betafunctions in the low scale regime are the
usual SM ones; in this paper we calculate the RG flow in both the mid
field and the high scale (inflationary) regime.

The idea of using the Higgs as the inflaton is an attractive one, not
least because it allows one to connect collider observables with
measurements of the early universe.  Since its inception the model
itself has come under a lot of scrutiny and criticism.  First,
unitarity is lost at high energies and the perturbative theory can
only be trusted for energies below the unitarity cutoff
\cite{burgess1,barbon,burgess2,Hertzberg1, cliffnew}.  Although it is uncertain
how to interpret this result as the cutoff is field dependent
(according to \cite{bezrukov4,linde_higgs,moss,He1}, all relevant physical
scales are always below the unitarity bound), it is clear that any new
physics living at this scale may affect the inflationary predictions
\cite{cliffnew,Hertzberg2}.  Second, given the currently measured central values
for the top and Higgs mass, the Higgs potential becomes unstable
around $10^{11}$ GeV, which would be disastrous for Higgs
inflation. However, the verdict is not yet out, as it only takes
$2-3\sigma$ deviations to push the instability bound all the way to
the Planck scale \cite{disc1, disc2, branchina,branchina2, archil, alexss} (in
the very recent note \cite{kniehl} absolute stability of the
electroweak vacuum is reported to be excluded by only 1.3 $\sigma$).

Even though these claims are still debated, and SM Higgs inflation may
still be alive, it is worth noting that constraints may be avoided in
modified set-ups with an extended Higgs sector.  Our results apply for
large non-minimal coupling, but apart from that they are equally
applicable to the various implementations of Higgs inflation.

The renormalization group equations (RGEs) in Higgs inflation have
been derived by several groups \cite{bezrukov_loop, bezrukov3,
  wilczek, barvinsky, barvinsky2, barvinsky3}, but they differ in
details.  The main source of disagreement comes from the choice of
frame, and the treatment of the Higgs sector (Does the Higgs decouple
from all fields? And the Goldstone bosons?). In previous work
\cite{damien, volpe} we have shown that the Jordan and Einstein frame
describe exactly the same physics, and that any difference stems from an
erroneous comparison of quantities defined in different frames. In
this work we will work in the Einstein frame.  Although dimensional
analysis indicates that some of the Goldstone boson (GB)-loop corrections are large, and
seem to spoil renormalization, gauge symmetry kicks in leading to
cancellations of these large contributions. We find Higgs inflation
is renormalizable in the effective field theory (EFT) sense \footnote{Higgs
  inflation is non-renormalizable as the field space metric and
  potential are non-polynomial. But this does not exclude that the
  theory is renormalizable in the EFT sense (as is the case in the
  IR). Our demands are that in the large and mid field regime the theory can
  be expanded in a small parameter $\delta$, and that all loop
  corrections can be absorbed in counterterms order by
  order. Truncating the theory at some finite order in $\delta$ gives
  a renormalizable EFT with a finite number of counterterms.}, and for
energies below the unitarity cutoff.

The small-field regime of Higgs inflation is where $\phi_0 \ll \mpl
/\xi$, with $\phi_0$ the value of the background Higgs field, $\xi$
the non-minimal Higgs-gravity coupling which is of order $10^4$ (well below experimental bounds \cite{Calmet,He2}), and
$\mpl$ the Planck mass.  In this regime the theory is effectively like
the SM and therefore renormalizable in the EFT sense. In the large
field regime ($\phi_0 \gg \mpl /\sqrt{\xi}$, corresponding to
inflation) the potential has an approximate shift symmetry, which
restricts the form of the loop corrections. As a result, all one-loop
corrections can be absorbed in the parameters of the classical theory,
and the EFT is renormalizable.  Somewhat surprisingly, we find the
same in the mid-field regime ($\mpl/\xi < \phi_0< \mpl/\sqrt{\xi}$),
even though it is far away from both an IR fixed point and the region in which the shift
symmetry applies.

In~\cite{damien} we have studied the renormalization of the
non-minimally coupled Higgs field in isolation, without any gauge or
fermion fields, and our findings were in line with the literature.
In this work we want to extend this previous analysis to the full SM. At first
glance this does not seem to be problematic. Due to the non-minimal
coupling to gravity, the coupling of the radial Higgs to both gauge field and
fermions is suppressed in the large field regime. One can simply
neglect all diagrams with these couplings. For example, loop diagrams
with a fermion or gauge boson loop always dominate over the
corresponding diagram with a Higgs loop. Effectively the Higgs
decouples from the theory.  However, the situation for the Goldstone
bosons (GBs) is more complex: their coupling to the gauge fields is also
suppressed, while the GB-fermion coupling is not. Upon going to
unitary gauge, this corresponds to a coupling of the fermion to the
longitudinal polarization of the gauge fields, and both the transverse
and longitudinal polarizations couple with the usual SM strength to
the fermions.  

All calculations are performed in the Einstein frame. For a discussion
of the equivalence of Einstein and Jordan frame, see
\cite{damien,volpe}. One of the main complications in the calculation
is that after transforming to the Einstein frame one ends up with
non-canonical kinetic terms for the Higgs and Goldstone field.  Due to
the nonzero curvature of the field space, it is impossible to make a
field transformation that brings the kinetic terms to their canonical
form.  Our approach here is to expand the action around a large
classical background value for the inflaton field, and use the
formalism of \cite{jinnouk, seery,Kaiser2} so that this background expansion
can be done maintaining covariance in the field space metric. 

In our calculation we have neglected the time-dependence of the
background field, as well as FLRW corrections and the backreaction
from gravity; we argue that these corrections are at most subleading. (The inclusion of gravity corrections to the Higgs part of the theory has been addressed, in a covariant way, in \cite{moss}.)
Moreover, we are neglecting higher order kinetic terms by evaluating the
field metric on the background. It would be an interesting but equally
challenging task to develop a framework that can get around this
latter limitation.

Our main results are the SM Higgs inflation RGEs in the three
regimes, where we included only the top-Yukawa coupling $y_t$.  We find that
Higgs/GB self-interactions and Higgs-fermion-interactions (but not
GB-fermion) can be neglected in the mid and large field regime;
Higgs/GB-gauge interactions decouple in the large field regime. This
gives the following betafunctions:
\begin{align}
(4\pi)^2 \beta_\lambda 
&= 24 \lambda^2\fac  + A
+(4\pi)^2 \cdot 4\lambda \gamma_\phi \nn \\
(4\pi)^2\gamma_\phi 
&= -\frac{\fac}{4} (3g_1^2+9 g_2^2) + 3y_t^2
\nn \\
(4\pi)^2\beta_{g_3} &= -7 g_3^3,\nn\\
(4\pi)^2\beta_{g_2} &= - \frac{(20 - \fac)}{6}g_2^3, \nn\\
(4\pi)^2\beta_{g_1} &=\frac{(40 + \fac )}{6}g_1^3 \nn \\
(4\pi)^2\beta_{y_t} &= \[\frac32 \fac y_t^3
-\( \frac{2}{3} g_1^2  +8g_3^2
\)y_t\] +(4\pi)^2 \cdot\gamma_\phi  y_t\nn\\
(4\pi)^2\beta_\xi\big|_{\rm mid, large} &= (4\pi)^2 \cdot2 \gamma_\phi \xi
\end{align}
with $A= (3/8)(2g_2^4 + (g_2^2+g_1^2)^2) -6y_t^4$ and 
% used \fac to be able to easily change notation
\be
\fac = 
\left\{
\begin{array}{ll}
1, & {\rm small} ,\\
0, &{\rm mid} , \; {\rm large}.
\end{array}
\right.
\label{fac}
\ee
These betafunctions break down at the boundary of the regimes, where
the EFT expansion in a small parameter is no longer valid; this gives
additional threshold corrections which we have not calculated.

All sign conventions used in this paper follow the QFT textbook by
Srednicki \cite{srednicki}, except for the sign of the Yukawa
interaction terms, which is opposite to Srednicki's.

%%%%%%%%%%%%%%%%%%%%%%%%%%%%%%%%%%%%%%%%%%%%%%%%%%%%%%
%%%%%%%%%%%%%%%%%%%%%%%%%%%%%%%%%%%%%%%%%%%%%%%%%%%%%%
%%%%%%%%%%%%%%%%%%%%%%%%%%%%%%%%%%%%%%%%%%%%%%%%%%%%%%

\section{Higgs inflation}
\label{s:hi}

In this section we give a brief overview of Higgs inflation and set
our notation.

\subsection{Lagrangian}

The Jordan frame Lagrangian\footnote{For considerations about initial conditions for Higgs inflation, and the possible inclusion of a $R^2$-term, see \cite{anupam}} is (using $-+++$ metric signature)
\be
\L^J = \sqrt{-g^J}\[ -\frac12 \mpl^2 \(1+ \frac{2\xi \Phi^\dagger
  \Phi}{\mpl^2} \) R[g^J] + \L^J_{\rm SM}\],
\ee
with 
\begin{align}
\L^J_{\rm SM} &= -\frac14 (f^a_{\mu\nu})^2
-\frac14 (F^a_{\mu\nu})^2 -\frac14 B_{\mu\nu}^2 
-(D_\mu \Phi)^\dagger (D^\mu \Phi) - \lambda(\Phi^\dagger \Phi - v^2/2)^2 \nn\\
&\hspace{0.45cm}+ \bar Q_L (i\slashed{D}) Q_L +\bar u_R (i\slashed{D})
u_R 
+\bar d_R (i\slashed{D}) d_R
- (y_d \bar Q_L \cdot \Phi d_R + y_u \bar u_R (i\sigma^2) \Phi^\dagger Q_L  +{\rm h.c} ),
\label{L_SM}
\end{align}
where $Q_L =(u \; d)^\top_L$. Further $f^a_\mn,F^a_\mn,B_\mn$ are the
SU(3), SU(2) and U(1) field strengths respectively.  The Higgs field
is SU(2) complex doublet, which we parameterize
\be
\Phi =
\frac1{\sqrt{2}} \( \begin{array}{c} \vp^+  \\
  \phi_0+\vp + i \theta_3 \end{array} \)  ,
\label{Hfields}
\ee
with $\phi_0$ the classical background, $\vp$ the Higgs field and
$\vp^+ = \theta_1+i\theta_2, \theta_3$ the GBs.
The covariant derivative acts on the Higgs field and fermions as
\begin{align}
D_\mu \Phi 
&= \(\partial_\mu -i g_2 A^a_\mu \tau^a - i Y_\phi g_1  B_\mu\)\Phi ,\nn \\
D_\mu Q_L &=(\partial_\mu -  i g_3 f^a_\mu t^a- i g_2
A^a_\mu \tau^a - i Y_Q g_1  B_\mu)Q_L, \nn \\
D_\mu u_R &=(\partial_\mu -  i g_3 f^a_\mu t^a
- i Y_u g_1  B_\mu)u_R,
\label{DH}
\end{align}
with $\tau^a = \sigma/2$ for the spinor representation. The
hypercharges are $Y_\phi =1/2$, $Y_Q = 1/6$ and $Y_u=2/3$.  At leading
order, the only fermion that matters to find the running of the SM
couplings is the top quark.

We reach the Einstein frame after a conformal transformation: $
g^E_{\mu\nu} = \Omega^2 g^J_{\mu\nu}$, with
\be
\Omega^2 = \(1+ \frac{2\xi \Phi^\dagger \Phi}{\mpl^2}\).
\label{Omega}
\ee
The Einstein frame Lagrangian becomes
\be
\L^E =\sqrt{-g^E}\[ -\frac12 \mpl^2 R[g^E] + \L_{\rm mat}^E\].
\ee
We neglect the expansion of the  universe, and take a Minkowski metric.

The gauge kinetic terms are conformally invariant.  The fermionic
kinetic terms can be made canonical via a rescaling  $\psi^E =
\psi/\Omega^{3/2}$; the net effect is then a rescaling of the Yukawa
interaction. All non-trivial effects of the non-minimal coupling are in the Higgs
sector.  
\begin{align}
\L^E_{\rm mat} 
 &= -\frac14 (f^a_{\mu\nu})^2
-\frac14 (F^a_{\mu\nu})^2 -\frac14 B_{\mu\nu}^2 
-\frac{1}{\Omega^2}(D_\mu \Phi)^\dagger (D^\mu \Phi) 
- \frac{3 \xi^2}{\mpl^2\Omega^4}
\partial_\mu (\Phi^\dagger \Phi) \partial^\mu (\Phi^\dagger \Phi)
\nn \\
&\hspace{0.45cm}+
 \bar Q^E_L (i\slashed{D}) Q^E_L +\bar u^E_R (i\slashed{D})
u_R^E 
+\bar d^E_R (i\slashed{D}) d^E_R
 \nn\\
&\hspace{.43cm}
- \frac{\lambda}{\Omega^4}(\Phi^\dagger \Phi - v^2/2)^2 
- (\frac{y_d}{\Omega} \bar Q^E_L \cdot \Phi d^E_R 
+ \frac{y_u}{\Omega}  \bar u^E_R (i\sigma^2) \Phi^\dagger Q^E_L  +{\rm h.c.} )
\end{align}
The Higgs kinetic term is non-minimal.  Let $\phi^i = \{\phi_R=\phi_0
+ \varphi,\theta_i\}$ run over the Higgs field and Goldstone
bosons. Then the metric in field space in component form is
\be
\L^E_{\rm mat} \supset-
\frac12 \gamma_{ij} \partial \phi_i \partial \phi_j = -\frac12 \[\frac{\delta_{ij}}{
  \Omega^2} + \frac{ 6
\xi^2}{\mpl^2\Omega^4} \chi_i \chi_j\] \partial \phi_i \partial \phi_j.
\label{non_can}
\ee
The curvature on field space $R[\gamma_{ij}] \neq 0$ (this point was made, and further generalized, in \cite{Kaiser1}, and the kinetic
terms cannot be diagonalized. At most one can diagonalize the
quadratic kinetic terms at one specific point in field space.

Consider the electroweak sector.  For the gauge bosons the kinetic
terms remain canonical in the Einstein frame. As far as the quadratic
action is concerned the action for the massive gauge bosons and
Goldstone bosons is simply three times the action of a U(1) theory.
To see this explicitly, consider the Higgs kinetic terms
\begin{align}
\L_{\rm higgs} &\supset 
-\frac{1}{\Omega^2}(D_\mu \Phi)^\dagger (D^\mu \Phi)  \nn \\
&=-\frac{1}{2\Omega^2}\[ \partial_\mu
\vp \partial^\mu \vp + \sum_{a=1}^3 (\partial_\mu \theta_a \partial^\mu
\theta_a -2g_a A^a_\mu (\phi \partial^\mu \theta_a
-\theta_a \partial^\mu \phi) + g_a^2 \phi^2A^a_\mu A_a^\mu) +...\].
%\label{Dphi}
\end{align}
The gauge boson mass eigenstates are $\{A_1,A_2, Z,A_\gamma\}$ with
\be
Z= \frac{1}{\sqrt{g_2^2 +g_1^2}}(g_2
A_3-g_1 B),\quad A_\gamma= \frac{1}{\sqrt{g_2^2 +g_1^2}}(g_1 A_3+g_2B) ,
\ee
and couplings
\be
g_a = \frac12 \times \left\{g_2,g_2,\sqrt{g_2^2 +g_1^2},0\right\}.
\label{ga}
\ee
This corresponds to three
massive and one massless field.  Note that we took the mass eigenstates
as real gauge fields, and used the real and imaginary parts of $W_+$,
rather than the complex states $W_\pm$.

From now on we will work in the Einstein frame. For convenience we
drop the superscript $E$, and work in Planck units $\mpl =1$.

%%%%%%%%%%%%%%%%%%%%%%%%%%%%%%%%%%%%%%%%%%%%%%%%%%%%%%%
\subsection{Three regimes}
\label{s:regimes}

Higgs inflation is non-renormalizable as the field space metric and
potential are non-polynomial. But this does not exclude that the
theory is renormalizable in the EFT sense over a limited field space.
Our demands are that in a given field regime the theory can be
expanded in a small parameter $\delta$, and that all loop corrections
can be absorbed in counterterms order by order. Truncating the theory
at some finite order in $\delta$ gives a renormalizable EFT with a
finite number of counterterms.  

\paragraph{Small field regime}
The small field regime corresponds to $\delta_s \equiv \xi \phi_0 \ll 1$. To
leading order in the expansion parameter $\delta_s$, the Lagrangian
reduces to the SM Lagrangian.

\paragraph{Mid field regime}
The mid field regime corresponds to $1/\xi < \phi_0 < 1/\sqrt{\xi}$.
In this regime we rescale $\xi \to \delta_m^{-2}\xi$ and $\phi_0 \to
\delta_m^{3/2} \phi_0$, such that both $\xi \phi_0^2 \propto \delta_m$
and $1/(\xi \phi_0)^2 \propto \delta_m$, and use $\delta_m$ as our
expansion parameter. (We should admit that formally this expansion can only be trusted in the middle of this regime.)

\paragraph{Large field regime}
Inflation takes place for field values $\delta_l \equiv 1/(\xi
\phi_0^2) \ll 1$.  The expansion in $\delta$ is equivalent to an
expansion in slow-roll parameters, since
$\eta=\mathcal{O}\left(\delta\right)$ and
$\eps=\mathcal{O}\left(\delta^2\right)$.

%%%%%%%%%%%%%%%%%%%%%%%%%%%%%%%%%%%%%%%%%%%%%%%%%%%%%%%%%%%%%%
%%%%%%%%%%%%%%%%%%%%%%%%%%%%%%%%%%%%%%%%%%%%%%%%%%%%%%%%%%%%%%
%%%%%%%%%%%%%%%%%%%%%%%%%%%%%%%%%%%%%%%%%%%%%%%%%%%%%%%%%%%%%%
\section{Covariant formalism and counterterms} \label{s:cov}

We want to investigate how the loop corrections and counterterms change in
the small, mid and large field regime. For simplicity, we first focus
on a U(1) Abelian Higgs model coupled to a left- and right-handed
fermion. The generalization to the full SM Higgs inflation is
postponed till sec. \ref{s:abelian_beta}. This way the effects of the non-minimal
coupling can be studied in a simple set-up, without all intricacies of
the chiral SM. Another advantage of the U(1) model is that gauge invariance and
the Ward identities assure that many counterterms are independent of
the gauge choice, which makes it easier to check the calculation.

%%%%%%%%%%%%%%%%%%%%%%%%%%%%%%%%%%%%%%%%%%%%%%%%%%%%%%%%%%%%%%
\subsection{Lagrangian in covariant fields} 

This subsection reviews the covariant formalism introduced in
\cite{jinnouk} and further worked out in \cite{seery,Kaiser2}. Given the curvature of
field space, it is very convenient to adopt an approach that maintains the
covariance of the equations. 

For a U(1) theory with a complex Higgs field and a left- and
right-handed Weyl fermion the Einstein frame matter Lagrangian is
\begin{align}
\L &= -\frac14 F_\mn F^\mn
-\frac12 \gamma_{ab} \partial_\mu \phi^a \partial^\mu \phi^b +i\bar \psi \slashed{\partial}\psi- V(\phi^a) -
\bar \psi F(\phi^a) \psi \nn\\
& \hspace{.43cm}
-g A (G^\theta \partial \phi-G^\phi \partial \theta) -\frac12 g^2 A^2
G
+ \(g q_L \bar \psi \slashed{A} P_L \psi +g q_R \bar \psi \slashed{A} P_R
\psi\),
\label{LU1}
\end{align}
with 
\begin{align}
V(\phi^a) &= \frac{\lambda}{4} \frac{|\phi_0+\varphi+i\theta |^4}{\Omega^4},\qquad
F(\phi^a) = \frac{y}{\sqrt{2}}\frac{\phi_0+ \varphi+
  i\gamma^5\theta}{\Omega},  \nn\\
G^\phi &= \frac{\phi}{\Omega^2},\qquad
G^\theta = \frac{\theta}{\Omega^2},\qquad
G = \frac{(\phi^2 + \theta^2)}{\Omega^2}.
\label{VF}
\end{align}

Now expand the Lagrangian around the background $\phi^a = (\phi_0(t)
+\varphi(x,t), \theta(x,t))$.  The fluctuation fields
$\delta\phi^a=(\varphi,\theta)$ are not in the tangent space at
$\phi_0^a$, and therefore do not transform as a tensor. We are led to
introduce the covariant fluctuation $Q^a = (h,\chi)$, which is related
to $\delta\phi^a$ via
\be
\phi^a = Q^a -\frac1{2!} \Gamma^a_{bc} Q^b Q^c +
\frac1{3!}\(\Gamma_{be}^a \Gamma^e_{cd} -\Gamma^a_{bc,d}\)
Q^b Q^c Q^d + ...
\label{Qdef}
\ee
This is the notation we will use throughout this paper:
$(\vp,\theta)$ are the fluctuations of the original Jordan frame
field (with $\phi_0$ the classical background field), and $(h,\chi)$
are the covariant fields.  Further we define the covariant time
derivative
\be
D_t = \frac{\dd \phi^a}{\dd t} \nabla_a. %= \dot \phi \nabla_\phi.
\ee
Note that in the limit $\dot \phi_0 =0$ this reduces to the usual
derivative $D_t = \partial_t$.

Now we can expand the action in covariant fluctuations. We neglect
FLRW corrections and the backreaction from gravity, as well as the
time-dependence of the background field $\phi_0$; we come back to this
in Sec.~\ref{s:checks}.  The result for the interaction Lagrangian is
\begin{align}
\L_{\rm int} &= -\ (V + V_{;a} Q^a +\frac1{2!} V_{;ab} Q^a Q^b +...)
-\bar\psi \ (F + F_{;a} Q^a +\frac1{2!} F_{;ab} Q^a Q^b +...) \psi 
\nn\\
& \hspace{.43cm}
-  g A \partial h ( G^\theta_{;a} Q^a +\frac1{2!} G^\theta_{;ab} Q^a
Q^b +...) 
+ g A \partial \chi ( G^\phi_{;a} Q^a +\frac1{2!} G^\phi_{;ab} Q^a Q^b +...)
\nn\\
& \hspace{.43cm}
-\frac12 g^2 A^2 ( G_{;a} Q^a +\frac1{2!} G_{;ab} Q^a Q^b +...)
+ \(g q_L \bar \psi \slashed{A} P_L \psi +g q_R \bar \psi \slashed{A} P_R
\psi\).
\label{L_cov}
\end{align}
All coefficients are evaluated on the background.  The subscript with
a semi-colon denotes the covariant derivative.  

We just found the Lagrangian for the covariant fields by Taylor expanding
using covariant derivatives.  An equivalent way of deriving the same
Lagrangian is solving the relation $\phi^i(Q^j)$ \eref{Qdef}
explicitly, substituting in the Lagrangian \eref{LU1}, 
\footnote{Equivalently, and probably more easily, the expansion $\phi^i(Q^i)$
 can be found by $\phi^i = \phi^i_{:a} Q^a + \frac1{2!} \phi^i_{:ab}
 Q^a Q^b+ ...$.}
and then Taylor expand in
the fields $Q^i$ (using partial derivatives). This point of view will
be useful when defining the counterterms in the next section.  Here
we just give the explicit form of \eref{Qdef} relating the original
Langragian fields $(\vp,\theta)$ to the covariant fields $(h,\chi)$:
\begin{align}
\vp &=  \(
h + \frac{ (h^2-\chi^2)}{2\phi_0} +  ..\) 
- \frac1\xi\( \frac{h^2}{\phi_0^3} + \frac{h^3}{3\phi_0^4} + .... \)
+ \frac{1}{\xi^2} \( \frac{h^2+\chi^2}{12\phi_0^3}  + .... \).\nn
\\
\theta & = \(\chi + \frac{h \chi}{\phi_0} +..\)
 - \frac1\xi\( \frac{h \chi}{\phi_0^3} +
  \frac{4h^2 \chi}{3\phi_0^4} + .... \)
+ \frac{1}{\xi^2} \( \frac{h \chi}{\phi_0^5} +
\frac{h^2 \chi}{12\phi_0^4} + .... \).
\label{Q_expl}
\end{align}
We checked that substituting this in the Lagrangian and expanding, we
indeed retrieve \eref{L_cov}.

%%%%%%%%%%%%%%%%%%%%%%%%%%%%%%%%%%%%%%%%%%%%%%%%%%%%%%%%%%%%%%
\subsection{Gauge fixing}

We have to add a gauge fixing and ghost Lagrangian, which can also be
expanded in covariant fields.

We fix the gauge via
\be
\L^E_{\rm GF}
=-\frac1{2\xi_G} \(\partial^\mu A_\mu - g G^\phi(\phi_0) \xi_G \theta
\)^2 .
%=-\frac1{2\xi_G} \(\partial^\mu A_\mu - 
%\frac{g \phi_0}{\Omega_0^2} \xi_G \theta\)^2
\label{L_GF}
\ee 
This removes the quadratic $A\partial \theta$ couplings from the
Lagrangian. In the small field regime $\Omega_0 \equiv \Omega(\phi_0)
=1$ and we retrieve the standard $R_\xi$-gauge.  We choose to write
the gauge fixing term in terms of the Jordan frame fields (as opposed
to the covariant fields) as these have
a well defined gauge transformation.

We work in Landau gauge $\xg = 0$. Then the ghost field decouples
\be
\L_{\rm FP}^E \big|_{\xg =0} =  -\partial_\mu \bar c \partial^\mu c.
\ee
%

%%%%%%%%%%%%%%%%%%%%%%%%%%%%%%%%%%%%%%%%%%%%%%%%%
\subsection{Feynman rules}
Now we can derive the Feynman rules from
the above action. First we define the effective couplings
\begin{align}
\L_{\rm int}& =
- \lambda_{m h n\chi} h^m \chi^n
- y_{m h n \chi}  h^m \chi^n \bar \psi
(i\gamma^5)^\alpha \psi 
- (g_{ A \partial h  mh n \chi } \partial h - g_{ A \partial
  \chi mh n\chi} \partial \chi ) A h^m \chi^n
\nn \\
& \hspace{.43cm}
-g_{2A mh n\chi} A^2 h^m \chi^n
+g_L \bar \psi \slashed{A} P_L \psi + g_R \bar \psi \slashed{A} P_R
\psi
\label{vertices1}
\end{align}
with $\alpha =1$ if the number $n= $ odd, and $\alpha=0$ otherwise
(signs are absorbed in the couplings).   All interactions are defined
with a minus sign (the only exception is for one of the derivative
interactions and the fermion-gauge interaction), and without numerical factors. This means that for a vertex
with $m$ $h$-fields and $n$ $\chi$-fields and with or
without fermion/gauge lines we have, respectively:
\begin{align}
V^{(m hn\chi)}&= (-i) m! n!  \lambda_{m hn\chi},\nn\\
V^{(m hn\chi2\psi)}&= (-i) m! n!  y_{m\phi
  n\chi}(i\gamma^5)^\alpha,\nn \\
V^{(m hn\chi2A)}&= (-i) 2! m! n!  g_{2Am hn\chi}.
\label{vertices2}
\end{align}
For the derivative interaction we get
\be
V^{(A \partial h m h n\chi)} =  -i g_{(A \partial h m h n\chi)} (-i
k^\mu),\quad
V^{(A \partial \chi m h n\chi)} =  i g_{(A \partial \chi m h n\chi)} (-i
k^\mu),
\ee
with $k$ the momentum running through the vertex.
The fermion, scalar and gauge propagators are given by:
\begin{align}
-i D_\psi(k) &= \frac{-i(-\slashed{k}+m_\psi)}{k^2+m_\psi^2-i\eps},
\nn \\
-i D_{Q^a}(k) &= \gamma^{aa} \frac{-i}{k^2 + (m^2)_a^a-i\eps},
\nn \\
-i D_{\mu\nu}(k)
&\stackrel{\xi_G = 0}{ =}-i \frac{g_{\mu\nu}-\frac{k_\mu k_\nu}{k^2}}{k^2 +m_A^2 -i\eps},
 \label{propagators}
\end{align}
with masses $m_\psi = F(\phi_0)$, $m_A = G(\phi_0)$, and $(m^2)^a_b =
\gamma^{ac} V_{;cb}(\phi_0)$. The scalar mass is diagonal, which we
used in the scalar propagator.  The propagators are standard except
for the metric factor in the scalar propagator.  The ghost field
decouples.

The above expansion of the Lagrangian, and corresponding Feynman rules
are equally valid in all field regimes (although in the small field
regime the notation is overkill). The explicit expressions for the
vertices are given in Appendix \ref{s:vertices}. The Higgs/GB mass
and self-interactions are suppressed in the mid and large field
regime, and we can neglect Higgs/GB loops in the quantum corrections.
Likewise, the interaction with the gauge field is suppressed in the
large field regime. The Higgs-fermion coupling is small in the mid and
large field regime, but the GB-fermion coupling is not: it has the
standard SM strength. This does not come as a surprise. The
gauge-fermion interactions are unaffected by the non-minimal
coupling. And since the GB is eaten to become the longitudinal
polarization of the gauge boson, it should have the same interaction
strength as the transverse polarizations.  A priori it is not clear
what the effects are on the betafunctions.  The explicit computation
is done in the next section.

%%%%%%%%%%%%%%%%%%%%%%%%%%%%%%%%%%%%%%%%%%%%%%%%%%%%
\subsection{Counterterms}

In this section we introduce counterterms. It proves convenient, or
maybe even necessary, to define the wave functions $Z_i$ in terms of the
original Jordan frame fields, rather than the covariant fields.  The
usual U(1) symmetry relations between the various $Z_i$-factors then
apply.

To define the counterterms we start with the Einstein frame Lagrangian
in terms of the {\it Jordan frame} fields \eref{LU1} and
rescale the bare fields (with label ``b") to physical fields (without label) via
\begin{align}
\phi_b &= \sqrt{ Z_\phi}  \phi, &
\theta_b &= \sqrt{ Z_\theta}  \theta, &
\psi_b &= \sqrt{Z_\psi}  \psi,&
A^\mu_b& = \sqrt{Z_A}  A^\mu,
\nn \\
\lambda_b &= Z_\lambda \lambda, &
\xi_b &= Z_\xi \xi, &
y_b &= Z_y y, &
g_b &= Z_g g,
\label{Zelement}
\end{align}
with $\phi = \phi_0 + \vp$. In Landau gauge $\xi_G=0$ the
wavefunctions are $Z_\vp =Z_{\phi_0} =Z_\theta$. 
We further define
\be
Z_i = 1+\delta_i.
\ee
We can then split the Lagrangian $\L =\L_{\rm renormalized} + \L_{\rm
  ct}$ with the counterterms proportional to $\delta_i$.  The total
Lagrangian is
\begin{align}
\L &= -\frac12  (Z_{g_{ab}}  g_{ab})  Z_\phi \partial \phi^a  \partial
\phi^b 
+ Z_\psi \bar \psi  i\slashed{\partial}\psi 
-\frac{1}{4}Z_A F_{\mu\nu} F^{\mu\nu} 
\nn \\ 
&
-  \frac{Z_\lambda Z_\phi^2 \lambda ( \phi^2 +
  \theta^2)^2}{ 4(1+ Z_\xi Z_\phi \xi \phi^2)^2}
- \frac{Z_\psi \sqrt{Z_\phi} Z_y  y}{\sqrt{2} (1+ Z_\xi Z_\phi \xi
  \phi_R^2)} \bar \psi (\phi + i\theta \gamma^5) \psi \nn\\
& -\frac1{ (1+ Z_\xi Z_\phi \xi
  \phi_R^2)} \( \sqrt{Z_A} Z_\phi Z_g g A (\phi \partial \theta
-\theta \partial \phi) + Z_A Z_\phi Z_g^2 \frac12 g^2 A^2
(\phi^2+\theta^2) \) 
\nn \\
&+Z_\psi  Z_g Z_A^{1/2} 
\(g q_L \bar \psi \slashed{A} P_L \psi +g q_R \bar \psi \slashed{A} P_R
\psi\).
\label{Lstart}
\end{align}
This can be simplified using the U(1) Ward identity
\be
Z_g = Z_A^{-1/2}.
\label{ward}
\ee
Moreover, as 
we will derive in the next section, we have in the mid and large field regime
(in the small field regime, $\xi$ drops out of the Lagrangian at
leading order, and no counterterm can be determined at this order)
\be Z_\xi = \frac1{Z_\phi}, \qquad {\rm for} \; \phi > \frac{\mpl}{\xi}.
\label{Z_xi}
\ee
This means $\Omega^2 = 1 + \xi \phi^2$ does not run (in the large
field regime it means $\delta = 1/(\xi \phi_0^2) \sim \e^{-h_0}$ does
not run). We can now use $\phi^i(Q^a)$ in \eref{Q_expl} to
rewrite the Lagrangian in terms of the covariant fields.  It is then
clear that all interactions arising from expanding the potential have
the same renormalization factor, namely $Z_\lambda Z_\phi^2$. Similar statements can be made for the gauge and Yukawa interactions.

We can also define ``composite'' wavefunctions for the interactions of
the covariant fields via
\begin{align}
\L 
\supset&= 
-  Z_{\lambda; mh n\chi} \lambda_{mh n\chi} h^m \chi^n
- Z_{y; m h n \chi}   y_{m h n \chi}  h^m 
\chi^n \bar \psi  (i\gamma^5)^\alpha \psi \nn \\
&
- Z_{2A; m h n \chi}   g_{2A m h n \chi} A^2  h^m \bar
\chi^n 
- Z_{A \partial Q^a; m h n \chi}   g_{A \partial Q^a m h n
  \chi} A \partial Q^a  h^m \bar
\chi^n .
\label{Lstart2}
\end{align}
It is straightforward to rewrite the composite wavefunctions in terms
of elementary ones \eref{Zelement}, by comparing terms in the two
actions above (\ref{Lstart},\ref{Lstart2}), when both written out in
covariant fields. As an explicit example, consider the potential in
the large field regime; using \eref{Lstart} it can be expanded as
\be
V = \frac{Z_\lambda}{ Z_\xi^2}  \frac{\lambda}{4\xi^2} 
+\frac{Z_\lambda}{Z_\phi Z_\xi^3} \lambda
\( - \frac{1}{\phi_0^4 \xi^3} h^2
- \frac{1}{3\phi_0^6 \xi^3} h^4
+ \frac{1}{12 \phi_0^6 \xi^5} \chi^2
- \frac{1}{108 \phi_0^{10} \xi^7} \chi^4
+ ...\)
\label{Zcompo}
\ee
%
%(* I think that the factor $1/3$ for (4h) should not be there... MP: i
%do get the $1/3$ factor for $\lambda_{4h} = 1/4! V_{:hhhh}$; you also
%have a factor 3 in your mathematica file*)
from which we read off 
\be
Z_{V_0} =  \frac{Z_\lambda}{ Z_\xi^2} ,\quad
Z_{\lambda 2h} = Z_{\lambda 4h} = Z_{\lambda 2\chi} = Z_{\lambda 4\chi} =
\frac{Z_\lambda}{Z_\phi Z_\xi^3} .
\label{Z_lambda}
\ee
Ignoring $\dot \phi_0$ corrections the Higgs kinetic terms are
\be
\L \supset -\frac12 Z_{2h} g_{h h} (\partial h)^2 
= -\frac12 Z_{2h} \frac{6}{\phi_0^2} (\partial h)^2
= -\frac12 \frac{Z_h}{Z_h} \frac{6}{\phi_0^2} (\partial h)^2.
\label{Z_kin1}
\ee
The leading counterterm vanishes.
The kinetic term for the GBs is
\be
\L \supset -\frac12 Z_{2\chi} g_{\chi\chi} (\partial \chi)^2 
= -\frac12 Z_{2\chi} \delta (\partial \chi)^2
= -\frac12 Z_h \delta (\partial \chi)^2.
\label{Z_kin2}
\ee
And thus 
\be
Z_{2h} -1 =\O(\delta), \qquad
Z_{2\chi} =  Z_\phi.
\ee
%
%(* is it confusing that counterterms are $\delta_i$ and expansion
%parameter is $\delta$?  should we change notation? *)

To derive these results we have used $Z_{h} = Z_{\chi} =Z_\phi$, i.e. the same
counterterms for the Jordan frame and covariant fields. As we will
discuss below, although this approximation is valid for the kinetic
terms, it is not in general.

%%%%%%%%%%%%%%%%%%%%%%%%%%%%%%%%%%%%%%%%%%%%%%%%%%%
\subsubsection{On the wavefunctions of the covariant fields}
\label{s:cov_wave}

Instead of defining the wavefunctions of the Jordan frame fields, we
could have tried to work with those of the covariant fields. Rather than beginning from \eref{Zelement}, we would then introduce $Z_{\phi_0}$, $Z_h$ and $Z_\chi$.  Consider the large field
regime. The potential is expanded in covariant fields as in
\eref{Zcompo}.  We can read off the counterterms from the parametric
dependence of the various terms. For the quadratic Higgs and GB interactions
we find $Z_h^{-1} Z_\xi^{-3}$ and $Z_h^{-2} Z_\xi^{-5}$ respectively, which should
be equal by gauge invariance.  This excludes setting $Z_{h} = Z_{\chi}
=Z_\phi$ in the potential, as it gives inconsistent
results.\footnote{One could leave the counterterms $Z_{h}$,
  $Z_{\chi}$, $Z_\phi$ unrelated a priori, and determine them by the
  requirement to absorb all (1-loop) divergencies. We tried this
  approach, and it fails. There is no consistent choice of counterterms
  that renders the theory finite.}

As we will now discuss, this approach breaks down for the GB
interactions.  We can understand where this stems from.  To derive the
$h^2$ and $h^4$ interactions at leading order, it is enough to only
keep the first term in the expansion \eref{Q_expl}.  Then taking the
relevant terms in \eref{Q_expl}, and setting $Q_b^i = \sqrt{Z_Q} Q^i$
we get
\begin{align}
\sqrt{Z_\phi} \vp &=  \sqrt{Z_{h}}  \(
h + \frac{ (h^2-\chi^2)}{2\phi_0} +  ..\) ,
\\
\sqrt{Z_\phi}  \theta & =\sqrt{Z_{\chi}}  \(\chi + \frac{h \chi}{\phi_0} +..\).
\end{align}
Thus for $h$ interactions we can take $Z_{h} = Z_{\chi} = Z_\phi$.
Similarly, for the kinetic terms, only the first order expansion is
needed, which is what we used above.

However, to derive the GB interactions, the leading and subleading
terms cancel, and to get the correct interaction one needs to expand
$\phi^i(Q^j)$ to sufficient high order in $\delta$.  In particular,
one needs also to take into account the last term in the expansion in
\eref{Q_expl}.  But then $Z_{h} = Z_{\chi} = Z_\phi =1/Z_\xi$ is no
longer a consistent solution; it would give the inconsistent relation
\be
\sqrt{Z_\phi} \vp =  \sqrt{Z_\phi}  (Q  + Q^2 + ...) +
Z_{\phi}^{3/2} ( Q^2+ Q^3 +...).
\ee
Thus for GB scattering one cannot really define $Z_{\chi}$, $Z_{h}$ in terms of the
elementary wavefunctions \eref{Zelement}.  Fortunately, this is also
not necessary, because we can simply use \eref{Zcompo}.

%%%%%%%%%%%%%%%%%%%%%%%%%%%%%%%%%%%%%%%%%%%%%%%%%%%%%%%%%%%%%%
\subsection{Approximations used} \label{s:checks}

Before diving into the calculation we first list here the
approximations made.

\begin{enumerate}
\item{We have dropped the time-dependence of the background field: $\dot \phi_0 \to 0$}
\item{We have neglected FLRW corrections and the backreaction of gravity}
\item{We have evaluated the field metric on the classical background.}
\end{enumerate}

1. In \cite{mp} we calculated the effective action in the SM regime, taking
into account the rolling of the classical background field
$\dot \phi_0$. Generalizing standard techniques to calculate the
effective action to the time-dependent situation, we found the
radiative corrections to both the classical potential and the kinetic
terms. This allowed us to extract both the $\delta_\lambda$ and
$\delta_\phi$ counterterms from the effective action.  We retrieved
the standard results. The time-dependence does not affect the form of
the counterterms.  Had we done the calculation in a time-independent
way, by neglecting $\dot \phi_0$, we would have found the same
$\delta_\lambda$ counterterm. 

In the large field regime the time-dependence enters also the kinetic
terms, which are non-minimal, and it may not be obvious that we can
neglect these effects.  However, the large field regime is the
inflationary regime, and all time-dependent corrections are slow roll
suppressed.  Working at leading order in the expansion parameter, as
we do, they can be neglected.

2. In \cite{damien,GMP} we calculated the effective action in the SM regime, in
a FLRW background.  We showed that when working in the {\it Einstein
  frame}, the backreaction from gravity can be neglected. The reason
is that the corrections are of the order of the slow roll parameter $\eps \sim
\delta^2$, which are small compared to $\eta \sim \delta$ and thus can
be neglected at leading order.

Doing the calculation in a FLRW background will give order $\mathcal{O}(H^2)$
corrections to the scalar masses, to the Higgs and GB mass in our
case. However, these masses only appear in diagrams with a Higgs and
GB in the loop, which thus also involve suppressed GB/Higgs
couplings. There is however one diagram that becomes of leading order
in the $\delta$-expansion, which is the last term of \eref{Pi_fermion}, giving the
GB loop correction to the fermion propagator. Nevertheless, this
diagram is still suppressed by $1/\xi$.  Hence, to be really sure that
FLRW corrections will not affect our results we have to work in the 
large $\xi \gg 1$ limit.

3. The kinetic terms for the GB/Higgs field are of the form
\be
\L \supset - \frac12 \gamma_{ij} \partial Q^i \partial Q^j =  -
\frac12 \gamma_{ij}(\phi_0) \partial Q^i \partial Q^j + ...
\ee
where  we have expanded the field space metric around the background. The
first term is quadratic and determines the structure of the
propagators. The higher order terms, denoted by the ellipses above, can
then be treated as additional interaction terms. It is hard to
systematically take into account the effects of  higher order
interactions, and we  have neglected them in our calculations in the
next section. Unfortunately, it seems that for at least one diagram
this is not a good approximation, as we discuss in section \ref{s:painful}.

%%%%%%%%%%%%%%%%%%%%%%%%%%%%%%%%%%%%%%%%%%%%%%%%%%%%%%%%%%%%%%
%%%%%%%%%%%%%%%%%%%%%%%%%%%%%%%%%%%%%%%%%%%%%%%%%%%%%%%%%%%%%%
%%%%%%%%%%%%%%%%%%%%%%%%%%%%%%%%%%%%%%%%%%%%%%%%%%%%%%%%%%%%%%

\section{One-loop corrections} \label{s:loop}

To derive the one-loop betafunctions for the gauge, Yukawa and Higgs
interactions, $g,y,\lambda$, it is enough to calculate the corrections
to the gauge, fermion and scalar propagator, which is what we will do
in this section. We will also compute corrections to 3 and 4-point
interactions. These will serve as consistency checks on the result,
which provide further checks on the validity of our approximations
(discussed in the previous section).  Another consistency check is the
comparison with the Coleman-Weinberg effective action \cite{CW}.

No field independent counterterms can be
defined for the whole regime, but it may be possible to define
renormalizable EFTs in the three different regimes. Then the hope is
that the threshold corrections in patching them together are small. To
find the result in a given regime, we plug in the explicit form of the
couplings expanded in the expansion parameter valid in this
regime. The expansion parameters were defined in subsection
\ref{s:regimes}; the explicit form of the couplings can be found in
Appendix \ref{s:vertices}.

%%%%%%%%%%%%%%%%%%%%%%%%%%%%%%%%%%%%%%%%%%%%%%%%%%%%%%%%%%%%%%
\subsection{Coleman-Weinberg effective action}

The Coleman-Weinberg calculation for a dynamical background field has
been performed in \cite{mp}. From this we can
extract $Z_{V_0}, Z_{2h}$, which should be consistent with the loop
corrections to the Higgs/GB propagator and self-scattering. The
effective action gets contributions from the bosonic, mixed and
fermionic loops respectively, and is for $\xg=0$ 
\begin{align}
\Gamma_{\rm CW} &= \frac{1}{32\pi^2\eps} \[
m_h^4 +m_\theta^4 +3 m_A^4  -4 m_f^4  + \frac{3}{2} m_{A\theta}^4
+ 4 m_f \ddot m_f  \] ,
\end{align}
with $m_{A\theta}^2 = -2 g \dot{\phi}_0/\Omega_0^2$. 
Adding classical and all one-loop contributions gives
\begin{align}
\Gamma & = 
 \frac12 \gamma_{hh} \dot{\phi}_0^2\[-Z_{2h} + \frac{1}{8\pi^2\eps} \( \frac{3 g^2}{1+ \xi \phi_0^2(1+6\xi)} - \frac{y^2}{\Omega_0^2 (1+
   \xi \phi_0^2(1+6\xi))} \)\]
\nn \\ &\hspace{0.45cm} +
\frac{\lambda \phi_0^4}{4\Omega_0^4}
\[ -Z_{V_0} +  \frac{1}{8\pi^2\eps} \(\lambda s(\phi_0) +3 \frac{g^4}{\lambda}
- \frac{y^4}{\lambda}\) \] ,
\label{Gamma}
\end{align}
%
%(* I get (see mathematica ``test section 4.1")
%\begin{align}
%\Gamma & = 
% \frac12 \gamma_{hh} \dot{\phi}_0^2\[-Z_{2h} + \frac{1}{32\pi^2\eps} \( \frac{12 g^2
%   }{1+ \xi \phi_0^2(1+6\xi)} - \frac{4y^2 \Omega_0^2}{ (1+
%   \xi \phi_0^2(1+6\xi))} \)\]
%\nn \\ &\hspace{0.45cm} +
%\frac{\lambda \phi_0^4}{4\Omega_0^4}
%\[ -Z_{V_0} +  \frac{1}{8\pi^2\eps} \(\lambda s(\phi_0) +3 \frac{g^4}{\lambda}
%- \frac{y^4}{\lambda}\) \] 
%\label{Gamma}
%\end{align}
%%
%MP: don't agree w/ fermion term.  you use $\dot m_f = y \dot
%\phi_0/(\sqrt{2} \Omega_0)$, but if you take derivative $\dot m_f
%= \partial_t \[ y \dot \phi_0/(\sqrt{2} \Omega_0)\]$ should also take
%into account the time dependence in $\Omega$.  *)
with
\be
s(\phi_0) = 
\frac{ \left(\phi_0^2 \xi  \left(1-2 \xi 
   \left(\phi_0^2 (6 \xi
   +1)-6\right)\right)+3\right)^2}{\Omega_0^4
   \left(\phi_0^2 \xi  (6 \xi +1)+1\right)^4}+\frac{ 1
   }{\Omega_0^4 \left(\phi_0^2 \xi  (6 \xi
   +1)+1\right)^2}.
\ee
It is clear that no field-independent counterterms can be defined over
the whole regime. Expanding the corrections in the respective regimes
we find (still applying the notation $Z_i=1+\delta_i$)
\be
\delta_{V_0}=
\frac1{8\pi^2\eps}\[10 \fac\lambda + 3\frac{g^4}{\lambda} - \frac{y^4}{\lambda}
\],
\label{dV0}
\ee
where we used notation \eref{fac}.  
Note that $\delta_{V_0} =
\delta_\lambda + 2\delta_\phi$ in the small and mid field regime, but
$\delta_{V_0} = \delta_\lambda -2\delta_\xi$ for large field.  As we
will see, consistency with Higgs/GB n-point functions requires
$\delta_\phi = -\delta_\xi$, as in \eref{Z_xi}, and thus
$\delta_{V_0}$ constrains the same elementary counterterms in the
whole regime.

Furthermore, we find
\be
\delta_{{2h}} = 
\frac1{8\pi^2\eps}\[\fac 3g^2 - \fac y^2
+\O(\delta) \].
\label{d2h}
\ee
In the  large field regime $\delta_{{2h}} = \O(\delta)$ is a
consistency check, but does not put any constraints on the elementary
counterterms \eref{Zelement}.  In the mid field regime we find
$\delta_{2h} =2(\delta_\phi+\delta_\xi) = \O(\delta)$, and thus to
lowest order \eref{Z_xi} is satisfied.  In the small field regime
$\delta_{{2h}} = \delta_\phi$, and we find an answer consistent with
\eref{Zphi} below.

%%%%%%%%%%%%%%%%%%%%%%%%%%%%%%%%%%%%%%%%%%%%%%%%%%%%%%%%%%%%%%
\subsection{Higgs/GB interactions} 

We start with the corrections to the Higgs propagator.  Compared to
the standard small-field calculation, the fermion loop is different because of the presence of
new fermion-Higgs/GB couplings. The gauge loop proceeds as in the
small field regime, with the only exception that the diagram with
derivative interactions cancels (at first order in the expansion parameter) as in the large field regime
$g_{A \partial h \chi} =- g_{A\partial \chi h}$ have opposite sign,
instead of being equal.  The result for the counterterm, fermion and
gauge, mixed gauge-GB and Higgs/GB loops is
\begin{align}
\Pi^{h} &= -\delta_{2h} \gamma_{hh}  k^2 -\delta_{\lambda_{2h}} 2\lambda_{2h}
+  \frac1{8\pi^2\eps} \bigg[
-
12 y_h^2 m_\psi^2 - 8 y_{2h} m_\psi^3 -2y_h^2 k^2
\nn \\
&\hspace{0.45cm} 
+6 g_{2A 2h} m_A^2 +6 g_{2A h}^2 +3 k^2 \gamma^{\chi\chi}\(\frac{g_{A \partial h \chi}+
g_{A\partial \chi h}}{2} \)^2
\nn \\
&\hspace{0.45cm} 
+12 \gamma^{hh}\lambda_{4h} m_h^2  +2 \gamma^{\chi\chi}
 \lambda_{2h2\chi} m_\chi^2 + 18 (\gamma^{hh})^2\lambda_{3h}^2 + 2 (\gamma^{\chi\chi})^2
 \lambda_{h2\chi}^2 
  \bigg] .
\label{Pi_h}
\end{align}
The $\gamma^{aa}$ factors stem from the Higgs and GB propagators.
Further, we used $m_h^2 = \gamma^{hh} (2\lambda_{2h})$ and similar for
the GB mass. This yields
\be
\delta_{\lambda_{2h}} = 
\frac1{8\pi^2\eps}\[10 \fac \lambda + 3\frac{g^4}{\lambda} - \frac{y^4}{\lambda}
 \],
\ee
while for the kinetic term we retrieve \eref{d2h}.

Comparing with the CW result we find that  $\delta_{\lambda_{2h}}
=\delta_{V_0}$. In the large field regime this gives the equality
$\delta_\lambda -\delta_\phi -3\delta_\xi =\delta_\lambda -2
\delta_\xi$, from which we get
\be
\delta_\xi = -\delta_\phi,
\label{delta_xi}
\ee
which assures that $\Omega$ does not run.   This is the same as
derived in the mid field regime from $\delta_{2h} = \O(\delta)$.
In the small field regime $\xi$ drops out of the Lagrangian at leading
order, and no relation for $\delta_\xi$ can be derived at this order.

The correction to the GB propagator is
\begin{align}
\Pi^{\chi} &= - \delta_{2\chi} \gamma_{\chi\chi} k^2 -\delta_{\lambda_{2\chi}} 2\lambda_{2\chi}
+  \frac1{8\pi^2\eps} \bigg[
- 4 y_\chi^2 m_\psi^2 - 8 y_{2\chi} m_\psi^3  -2y_\chi^2 k^2
\nn \\
&\hspace{0.45cm} 
+6 g_{2A 2\chi} m_A^2 +3 k^2 \gamma^{hh}\(\frac{g_{A \partial h \chi}+
g_{A\partial \chi h}}{2} \)^2
\nn \\
&\hspace{0.45cm} 
+12 \gamma^{\chi\chi}\lambda_{4\chi} m_\chi^2  +2 \gamma^{hh}
 \lambda_{2h2\chi} m_h^2 + 4 \gamma^{hh}\gamma^{\chi\chi}
 \lambda_{h2\chi}^2 
  \bigg] .
\label{Pi_chi}
\end{align}
We find $\delta_{\lambda_{2h}} =\delta_{\lambda_{2\chi}}$ in all three
regimes, as required by gauge invariance. Further we have
\be
\delta_{2\chi}  = \delta_\phi=
\frac1{8\pi^2\eps}\[ 3\fac g^2 -y^2\].
\label{Zphi}
\ee

It is interesting to note that in the large field regime the
counterterm $\delta_{\lambda 2\chi} = \O(\delta^3)$ whereas the
individual fermion and gauge loop diagrams in \eref{Pi_chi} are $\O(\delta^2)$.
Renormalizability thus requires the two fermion diagrams to cancel at
leading order, to
end up with an $\O(\delta^3)$ loop correction. This is indeed what
happens.  This intricate cancellation is even more pronounced when we
consider corrections to higher $n$-point GB scattering.  For example, the
structure of the fermion contribution to the four-point GB vertex is
\begin{align}
V^{(4\chi)} &= -\delta_{\lambda 4\chi} 4! \lambda_{4\chi}
\nn \\ &\hspace{0.45cm}
+ \frac{ 4!} {8\pi^2\eps} \Bigg[
36 \lambda_{4\chi}^2(\gamma^{\chi\chi})^2 + \lambda_{2h2\chi}^2(\ghh)^2
-m_h^2 \lambda_{2h4\chi} \ghh -15 m_\chi^2 \lambda_{6\chi}\gamma^{\chi\chi} + 8\lambda_{h4\chi} \lambda_{h2\chi} \ghh \gamma^{\chi\chi}
\nn \\
& \hspace{1 cm} - 4  y_{4\chi} m_\psi^3 -  4y_{3\chi}
y_\chi (m_\psi^2+\frac12 k^2)- 6 y_{2\chi}^2
(m_\psi^2+\frac16 k^2) -4 y_{2\chi}
y_\chi^2 m_\psi - y_\chi^4 
\nn \\
& \hspace{1 cm} + 3(g_{2A2\chi}^2 + g_{2A4\chi} m_A^2) \Bigg] .
\label{4theta}
\end{align}
Now the counterterm on the first line is
$\delta_{\lambda 4\chi} = \O(\delta^5)$. The GB and Higgs loop
diagrams on the second line above give $\O(\delta^6)$
corrections and can be neglected.  All individual fermion loop
diagrams --- the terms on the third line --- and all individual gauge
loop diagrams --- the terms on the fourth line --- are $\O(\delta^3)$,
much larger than the counterterm.  Thus both the leading and
subleading contributions need to cancel when adding the diagrams.
This is indeed what happens and we find
$\delta_{\lambda_{2h}} = \delta_{\lambda_{4\chi}}$ as required by
gauge invariance.  This intricate cancellation, and the need to go to
sub-sub-leading order in the $\delta$-expansion, is the reason we
cannot easily define the wavefunctions for the covariant fields, as
discussed in section \eref{s:cov_wave}.

Note, however, that the $k^2$-term in \eref{4theta} above does not cancel,
and gives a correction that cannot be absorbed.~\footnote{Here
  $k$ is the momentum flowing in the loop at one of the vertices; the propagator structure is  $\int \dd^4 l D_\psi(l) D_\psi(l+k)$.}  For this we have to add a new dimension-6 counterterm which is
a four-point $\chi$-interaction with two derivatives; very schematically
\be
\L \supset \frac{1}{\Lambda^2} \chi^2 (\nabla \chi)^2, 
\ee
with a cutoff 
\be
\Lambda \sim (y_{3\chi}y_\chi +\frac12 y_{2\chi}^2)^{-1/2} 
\gtrsim \Lambda_{\rm unitarity} \label{CU1}
\ee
that is equal to or larger than the unitarity cutoff
\be
\Lambda_{\rm unitarity} \sim 
\left\{\frac1{ \xi} ,
\phi_0 ,
\frac{1}{\sqrt{\xi}}\right\}
\label{unitarity}
\ee
in the small, mid and large field regime regime.
The EFT breaks down
for energy scales beyond the unitarity cutoff. The new
counterterms needed to absorb divergencies enter at even higher
scales. As such they do not put further constraints on the domain of
validity of the EFT.

%%%%%%%%%%%%%%%%%%%%%%%%%%%%%%%%%%%%%%%%%%%%%%%%%%%%%%%%%%%%%%
\subsection{Yukawa interactions}

We first calculate the corrections to the fermion propagator. The
fermion-gauge coupling is standard over the whole field range, and the
gauge loop gives the same result in all three regimes. This is not the case
for the Higgs and GB loop, as the former is suppressed in the mid and large
field regime.
The fermion two-point function is
\begin{align}
\Pi^{(2\psi)} &=  -\delta_\psi \slashed{k} - \delta_{m_\psi} m_\psi
+ \frac1{8\pi^2\eps} \bigg[ -3 m_\psi g^2 q_L q_R
\nn \\ 
&\hspace{0.45cm} 
+y_h^2 \gamma^{h h}
(m_\psi -\frac12 \slashed{k}) -y_\chi^2 \gamma^{\chi\chi}
(m_\psi +\frac12 \slashed{k}) +y_{2h}\gamma^{h h} m_h^2 +y_{2\chi}\gamma^{\chi\chi}
m_\chi^2 \bigg],
\label{Pi_fermion}
\end{align}
where we used $g_{A\bar \psi_{L,R} \psi_{L,R}} = g q_{L,R}$ in all
three regimes.  There is a minus sign difference between the two
$\slashed{k}$ terms, which originates from the $(i\gamma^5)$ in
vertices with an odd number of GBs. This gives for the counterterms
\begin{align}
\delta_\psi &= -\frac1{8\pi^2\eps} \[\frac{y^2}{4}(\fac+1) \],
\nn \\
\delta_{m_\psi}&=\delta_y +\delta_\psi+ \frac12 \delta_\phi
=\frac1{8\pi^2\eps}\(-3 g^2 q_L q_R + \frac12(\fac-1) y^2\) .
\label{Z_2psi}
\end{align}
The GB and Higgs contribution add in $\delta_\psi$ and cancel in
$\delta_{m_\psi}$ in the small field regime; in the mid and large
field regime only the GB contribution survives.

The Yukawa interactions, both Higgs-fermion and GB-fermion,
should give consistent results.  Indeed we find
\begin{align}
V^{\Psi \bar{\Psi}h}_{\rm tot} &=- \frac{1}{8\pi^2\eps}\Bigl(
-y_h^3\ghh
+y_h y_\chi^2\gamma^{\chi \chi}
-3m_h^2 y_{3h}\ghh
-m_\chi^2 y_{h2\chi}\gamma^{\chi \chi}
-2y_h y_{2h}\ghh \left(\slashed{k}+2m_\Psi\right)
\nn\\ & \quad
 -\frac{y_\chi y_{h\chi}\gamma^{\chi \chi}}{2} \left(\slashed{k}-2m_\Psi\right) 
 -6 y_{2h} \lambda_{3h} (\ghh)^2
-2 y_{2\chi} \lambda_{h2\chi}(\gamma^{\chi \chi})^2
+ 3 q_L q_R g_{\bar\Psi A\Psi}^2 y_h\Bigr) 
-\delta_{y_h}y_h,
\end{align}
and
\begin{align}
V^{\Psi \bar{\Psi}\chi}_{\rm tot} &=- \frac{i \gamma^5}{8\pi^2\eps}\Bigg( 
y_h^2 y_\chi \ghh
- y_\chi^3\gamma^{\chi \chi}
 -m_h^2 y_{2h\chi}\ghh
-3  m_\chi^2 y_{3\chi}\gamma^{\chi \chi}
 -\frac{y_h y_{h\chi}\ghh}{2} \left(\slashed{k}+2m_\Psi\right) \nn\\
 & \quad 
-2y_\chi y_{2\chi}\gamma^{\chi \chi} \left(\slashed{k}+2m_\Psi\right)  
- 2y_{h\chi}   \lambda_{h2\chi}\ghh\gamma^{\chi \chi}
 +3q_L q_R g_{\bar\Psi A\Psi}^2 y_\chi
\Biggr) 
- i\gamma_5\delta_{y_\chi} y_\chi. 
\end{align}
This indeed gives $\delta_{y_h} = \delta_{y_\chi} = \delta
_{m_\psi}$, with the latter given in \eref{Z_2psi}.

However, once again there is a small glitch as the $\slashed{k}$ terms in
both expressions do not cancel.  New non-renormalizable counterterms
need to be added, schematically of the form
\be
\L  \supset\frac{1}{\Lambda} \bar\psi  (h + i\gamma^5 \chi)\slashed{\partial} \psi 
\ee
with cutoff 
\be
\Lambda \sim (\gamma^{\chi\chi}y_\chi y_{2\chi})^{-1}
\gtrsim \Lambda_{\rm unitarity}. \label{CU2}
\ee
Since the cutoff exceeds the unitarity cutoff, these terms do no
affect the range of validity of the EFT in the three regimes.

%%%%%%%%%%%%%%%%%%%%%%%%%%%%%%%%%%%%%%%%%%%%%%%%%%
\subsection{Gauge interactions}

We begin with the gauge boson propagator.
\begin{align}
\Pi^A_\mn &= -\delta_A( k^2 g_{\mu\nu} -  k_\mu k_\nu) 
-\delta_{m_A} m_A^2 g_\mn \nn \\
&\hspace{0.45cm}
+ \frac1{8\pi^2 \eps}\Bigg[ \(3 \gamma^{hh} g_{h2A}^2 +
2 \gamma^{hh} g_{2h2A} m_h^2 
+ 2\gamma^{\chi\chi} g_{2\chi 2A} m_\chi^2\) g^\mn\nn \\
&\hspace{0.45cm}+
\gamma^{hh} \gamma^{\chi \chi} \[-\frac14 (g_{A\partial h \chi}+g_{A\partial \chi h})^2 \(\frac{k^2}{3} +m_h^2
+m_\theta^2\) g^\mn + (g_{A\partial h \chi}^2-g_{A\partial h \chi}g_{A\partial \chi h}+g_{A\partial \chi h}^2)\frac13k^\mu k^\nu\]\nn \\
&\hspace{0.45cm}
-\frac23 (k^2 g^\mn -k^\mu k^\nu)\( g_L^2 +g_R^2\)
-2 (g_L -g_R)^2 m_\psi^2 \Bigg]
\end{align}
It should be remembered that we normalized $q_\phi =1$. The
counterterms are
\be
\delta_A = -\frac1{8\pi^2 \eps} g^2\( \fac \frac13 +\frac23 q_L^2
+\frac23 q_R^2\).
\ee
Using the Ward identity $2\delta_g= -\delta_A $ \eref{ward}, it
follows that
$\delta_{m_A} = 2\delta_g + \delta_\phi + \delta_A =
\delta_\phi$.
Reading off $\delta_{m_A} $ from the above expression, and comparing
with our earlier result \eref{Zphi} for $\delta_\phi$, we indeed find agreement.

In the large field regime there is also a derivative interaction at leading
order that is not transversal and cannot be absorbed in $\delta_A$. We
find a term
\be
\Pi^A \supset \frac1{8\pi^2 \eps} \frac{g^2}{(1+6\xi)} k^\mu k^\mu.
\label{transversal}
\ee
This term can be neglected only for $\xi \gg 1$.  This is the only
place where this extra condition is needed.  The transverse term breaks
the Ward identities in the Landau gauge, and should not be there.  It
arises as a consequence of our approximations, discussed in more
detail in section \ref{s:painful}. We  are not too worried about this
term, as it is absent in the large $\xi$ limit. But moreover, it is
also a gauge dependent term.  We could have chosen a gauge fixing 
\be
\L^E_{\rm GF}
=-\frac1{2\xi_G} \(\partial^\mu A_\mu - g \frac{\phi_0}{\Omega_0} \xi_G \chi
\)^2 
\ee 
defined in terms of the covariant fields rather than the Jordan frame
fields \eref{L_GF}. In Landau gauge, this gauge fixing gives the same
results for all other diagrams, but now also the transversal part
\eref{transversal} vanishes.

As a consistency test we also calculated the $2A2h$ interaction, which
gives
% [check, many different conventions
%$\lambda_{4h} \to lambda_{4h} /4,\lambda_{2h2\theta} \to
%lambda_{2h2\theta} /2, g^2_{2h 2A} \to g_{2h2A}/2, y_i \to
%y_i/\sqrt{2}$]
%
\begin{align}
V^{2A2h} &=\frac{g^{\mu\nu}}{8\pi^2\eps}
\Bigl(
48  \lambda_{4h} g_{2A2h}q_\Phi^2(\ghh)^2 
+8 \lambda_{2h2\chi} g_{2A2\chi} (\gamma^{\chi\chi})^2   
-   (g_{A\partial h \chi } +g_{A\partial \chi h })^2\ghh
                    (\gamma^{\chi\chi})^2 \lambda_{2h2\chi} 
\nn\\
& \qquad
- 6 (g_{A\partial h \chi } +g_{A\partial \chi h })^2(\ghh)^2
  \gamma^{\chi\chi} \lambda_{4h} \
 +24 g_{2A2h}^2  \ghh 
- 4 (q_L-q_R)^2y_h^2 g_{\bar \Psi A \Psi}^2 
\nn\\
& \qquad
-8 (q_L-q_R)^2 y_{2h}g_{\bar \Psi A\Psi}^2 m_\Psi
\Bigr)
-4 \delta_{g_{2A2h}} g_{2A2h}  g^{\mu\nu},
\end{align}
yielding $\delta_{m_A} = \delta_{g_{2A2h}}$, as it should.

%%%%%%%%%%%%%%%%%%%%%%%%%%%%%%%%%%%%%%%%%%%%%%%%%
\subsection{Gauge-fermion vertex} \label{s:painful}

There is one interaction that does not give a consistent result, which
is the fermion-gauge coupling.  To calculate it the important terms in
the Lagrangian are
\be
\L = -\frac{3\xi^2}{\Omega^4} \partial_\mu (\Phi^\dag
\Phi) \partial^\mu (\Phi^\dag \Phi)
-\frac{ \left(D_\mu \Phi\right)^\dag D^\mu \Phi}{\Omega^2}
+  \sum_{i=L,R}i\bar \Psi_i \slashed{D} \Psi_i
-  \frac{y  \Phi \left(\bar \Psi_L \Psi_R+{\rm
      h.c.}\right) }{\Omega} .
\label{oerlag}
\ee
There are three one-loop diagrams: 1) a GB loop with the photon
attached to the fermion line, 2) a Higgs loop with the photon attached
to the fermion line, and 3) a mixed Higgs-GB loop with the photon
attached via a derivative interaction to $\Phi$.  The result is
\begin{align}
V_{\rm loop}^{ \bar \Psi A_\mu \Psi} &= -\frac{\gamma^\mu}{16\pi^2 \eps} \Bigg[
g_{\bar \Psi A \Psi} (y_\chi^2 \gamma^{\chi\chi} +y_h^2 \ghh)
\left(q_L  P_R+q_R P_L\right) 
 \nn \\
& \qquad + \left(g_{A\partial h \chi }+g_{A\partial \chi h }\right) y_h
y_\chi \ghh \gamma^{\chi\chi}
 \left(q_\Phi P_L - q_\Phi P_R\right) \Bigg], \nn \\
V_{\rm CT}^{ \bar \Psi A_\mu \Psi} &= -\delta_{ \bar \Psi A_\mu \Psi}g_{\bar \Psi A)\mu \Psi} \gamma^\mu(q_L P_L + q_R P_R) .
\end{align}
In the small field regime this reduces to
\be
V_{\rm loop}^{ \bar \Psi A_\mu \Psi} =-\frac{g y^2 \gamma^\mu}{16\pi^2}\frac{1}{\eps}\left(q_L P_L + q_R P_R  \right),
\ee
where we used gauge invariance: $ q_\Phi - q_L + q_R =0$. This
expression can be absorbed in the counterterm, which is proportional
to $\left(q_L P_L + q_R P_R \right)$ as well.  In the large field regime,
however, the diagrams with a Higgs loop are suppressed and we get
\be
V_{\rm loop}^{ \bar \Psi A_\mu \Psi} = \frac{\gamma^\mu}{32\pi^2}\frac{1}{\eps} \Biggl[  \left( \frac{gy^2}{6\xi^3 \phi_0^4} + gy^2   \right)\left(q_L P_R+q_R P_L\right) +\frac{g y^2}{3\xi^3 \phi_0^4}\left(q_\Phi P_L - q_\Phi P_R\right)          \Biggr].
\ee
We cannot combine the two parts, and will not get something
proportional to $\left(q_L P_L + q_R P_R \right)$. The same problem
arises in the mid-field regime.

This result in the large field regime breaks gauge invariance
explicitly. How did it arise? When we repeat the calculation without the
first term in \eref{oerlag}, the Higgs and GB field still have the
same propagator. As a result all three diagrams contribute and the
result adds up to something gauge invariant.  However, when we include
the first term, things go wrong as the Higgs propagator is now
suppressed compared to the GB propagator.  Note however, that the
first term is explicitly gauge invariant. It is our approximation that
breaks the gauge invariance, when we evaluate the metric on the
background $\gamma_{ij} (\phi,\theta) = \gamma_{ij} (\phi_0)$.  In
particular, for the first term we set
\be
\L \supset -\frac{3\xi^2}{\Omega^4} \partial_\mu (\Phi^\dag
\Phi) \partial^\mu (\Phi^\dag \Phi) = -\frac{3\xi^2
  \phi_0^2}{2\Omega_0^4} (\partial_\mu \phi)^2+ ...
\ee
where the ellipses denote neglected higher order derivative
interactions (to be precise: higher n-point interactions with two
derivatives).  We listed this as the third approximation in subsection
\ref{s:checks}. These higher order terms need to be included to obtain
a gauge invariant result.  Unfortunately, it does not seem
straightforward to do so. We would like to postpone the setup of a
framework able to handle higher order derivative terms to future work,
leaving a loose thread to our current calculation. However, since the
calculation of the two-point function involves lower order vertices, we
expect it to be less prone to our approximation.

\section{RGE equations} \label{s:beta}

First we give the betafunctions for the Abelian-Higgs model with a
non-minimal coupling, then in subsection \ref{s:SM_beta} we generalize
to full SM Higgs inflation.

%%%%%%%%%%%%%%%%%%%%%%%%%%%%%%%%%%%%%%%%%%%%%%%%%
\subsection{Abelian Higgs model} \label{s:abelian_beta}

First we list all the counterterms, found in the previous section:
\begin{align}
\delta_\phi =-\delta_\xi& =\frac1{8\pi^2\eps} \(3g^2 \fac - y^2\),\nn \\
\delta_{\lambda_{2h}} = \delta_{\lambda_{4h}}=\delta_{\lambda_{2\chi}} = \delta_{\lambda_{4\chi}}&=\frac1{8\pi^2\eps}\(
10 \lambda \fac  +3 \frac{g^4}{\lambda} -\frac{y^4}{\lambda} \) ,\nn\\
\delta_\psi &=  
-\frac{1}{8\pi^2\eps}\left(\frac{y^2}{4} (\fac+1)  \right),\nn\\
\delta_{m_\psi} =  \delta_{y_{h\bar \psi \psi}} =\delta_{y_{\chi\bar
    \psi \psi}} &
=\frac1{8\pi^2\eps}\(-3 g^2 q_L q_R+\frac12 y^2 (\fac-1) \) ,\nn\\
\delta_A &= -\frac{1}{8\pi^2\eps}g^2 (\frac{1}{3}q_\phi^2 \fac
+\frac{2}{3}q_L^2+\frac{2}{3}q_R^2).
 \nn\\
\end{align}
The counterterms for the couplings in the Lagrangian are then given by
$\delta_\lambda=\delta_{\lambda_{4h}} -2 \delta_\phi$,
$\delta_y = \delta_{m_\psi} -1/2 \delta_\phi -\delta_\psi$, and the
Ward identiy $\delta_g =-1/2 \delta_A$ respectively. From this the
beta-functions can be found via
$\beta_\lambda = \lambda (\eps \delta_\lambda)$, and similarly for the
Yukawa, gauge and non-minimal Higgs-gravity coupling.  This gives
\begin{align}
 \beta_\lambda & = \frac1{8\pi^2}\( 10 \fac \lambda^2 +3
g^4 -y^4 -6 \fac g^2 \lambda +2 y^2\lambda \) ,\nn \\
\beta_y & = \frac1{8\pi^2}\(-3 q_L q_R g^2 y-\frac32 \fac (q_L-q_R)^2g^2y +\frac14(1+3\fac) y^3\)\nn \\
\beta_g&=\frac{1}{8\pi^2} g^3 \(\fac\frac{1}{6}q_\phi^2
+\frac{1}{3}q_L^2+\frac{1}{3}q_R^2\) ,\nn \\
\beta_\xi \big|_{\rm mid, large} &=-\frac1{8\pi^2}  (\fac 3 g^2-y^2) \xi.
\label{betaU1}
\end{align}
We have not derived $\beta_\xi$ in the small field regime.  At leading
order all $\xi$ dependence drops out of the Lagrangian in the SM
regime. Since our computation relies on the approximations listed in section \ref{s:checks}, which fail beyond the leading order, we have to leave the computation of $\beta_\xi$ in the small field regime open.

%%%%%%%%%%%%%%%%%%%%%%%%%%%%%%%%%%%%%%%%%%%%%%%%%%%%%
\subsection{SM Higgs inflation} \label{s:SM_beta}

Our results for a U(1) theory can be extended to the full Standard
Model (SM) Higgs inflation.  Working in background field gauge~\footnote{In
  practice, it is not so easy to calculate diagrams in the background
  field gauge, as it is unclear how to expand the Lagrangian in
  covariant fields with a shifted metric.}, the symmetries of the
classical effective action are similar to those of the U(1) theory.
The main difference is that now there are 3 GBs, the top quark has
three colors, and one needs to sum over the strong, weak and
hypercharge interactions.

\paragraph{Higgs coupling} First we extend the U(1) results in the small
field regime to the full SM beta-functions, which can be found for
example in \cite{sher}. The SM betafunction for the Higgs self-coupling
is a straightforward generalization of the U(1) result:
\begin{align}
\beta_\lambda &= \frac{1}{8\pi^2} \[(9+n_\theta) \lambda^2  +3 \sum_a
g_a^4 - n_c  y_t^4
- 2\lambda  ( 3\sum_a g_a^2
-n_c  y_t^2) \] \nn \\
 &= \frac{1}{8\pi^2}  \[12 \lambda^2  + \frac{3}{16}
\(2 g_2^4 + (g_2^2+g_1^2)^2\) -3 y_t^4
- 2\lambda  (\frac94 g_2^2 +\frac34 g_1^2 -3y_t^2) \] ,
\end{align}
with $n_\theta =3$ the number of GBs, $n_c =3$ the number of colors
and the $g_a$ are given in \eref{ga}.  We have only included the running of the
top Yukawa.  Generalizing from the U(1) model, it follows that the
$\lambda^2$ and the $\lambda g_{1,2}^2 $ terms are suppressed in the
mid and large field regime. This gives
\be
\beta_\lambda 
= \frac{1}{(4\pi)^2}  \[24 \lambda^2\fac  + \frac{3}{8}
\(2 g_2^4 + (g_2^2+g_1^2)^2\) -6 y_t^4
- \lambda  (9 g_2^2 +3{g_1}^2)\fac  +12 y_t^2 \lambda) \].
\ee

\paragraph{Gauge coupling} 
For an $SU(N)$ group the betafunction is 
%(see for example p74 of
%\url{http://www.nikhef.nl/~t58/BSM.pdf} )
%
\be
\beta(g) \bigg|_{SU(N)}= -\frac{g^3}{96\pi^2} \( 22N - 2 n_f - n_H\),
\ee
with $n_f$ the number of Weyl fermions and $n_s$ the number of complex
Higgs fields, both in the fundamental representation.
For an Abelian group there is no contribution from the gauge field,
and the formula becomes
\be
\beta(g) \bigg|_{U(N)}= \frac{g^3}{96\pi^2} \(4 \sum q_f^2 +\sum q_s^2\),
\ee
with $q_f,q_s$ the charges of the Weyl fermion and real scalars
respectively. This reproduces our result in the small field regime.
In the mid and large field the Higgs and GB contributions to the running are
absent; this does not affect QCD, but for the EW sector we get
\begin{align}
\beta_{g_3} &= -\frac{7}{(4\pi)^2} g_3^3,\nn\\
\beta_{g_2} &= -\frac{1}{(4\pi)^2}  \frac{(20 - \fac)}{6}g_2^3, \nn\\
\beta_{g_1} &= \frac{1}{(4\pi)^2} \frac{(40 + \fac)}{6}g_1^3. \nn
\end{align}

%(* remove this part about counting in SM *) 
%
%For SU(3) there are $6 \times 2$ colored Weyl
%fermion (6 quarks). For SU(2) there are $4\times 3$ doublets (4 per
%generation) and one complex Higgs doublet. This gives
%%
%\begin{align}
%\beta(g_3) &=  -\frac{g_3^3}{96\pi^2} \( 11 \times 6 - 2 \times 12\)
%=-\frac{7g_3^3}{(4\pi)^2} 
%\nn \\
%\beta(g_2) &=  -\frac{g_2^3}{96\pi^2} \( 11 \times 4 - 2 \times 12 - 1\)
%=-\frac{19g^3}{6 (4\pi)^2} 
%\end{align}
%%
%For an Abelian group there is no contribution from the gauge field,
%and the formula becomes
%%
%\be
%\beta(g) \bigg|_{U(N)}= \frac{g^3}{96\pi^2} \(4 \sum q_f^2 +\sum q_s^2\)
%\ee
%%
%with $q_f,q_s$ the charges of the weyl fermion and real scalars
%respectively. This reproduces our result in the small field
%regime. For the SM hypercharge, the fermion contribution for a single
%family is
%%
%\be
%4\sum q_f^2 = 4 \times\[ 6 \times (\frac16)^2 +3 \times (\frac13)^2 + 3
%\times (-\frac23)^2 +2 \times (-\frac12)^2 + 1 \] = \frac{40}{3}
%\ee
%%
%The scalar contribution comes from the complex Higgs doublet:
%%
%\be
%\sum q_s^2 = 4 \times (\frac12)^2 =1.
%\ee
%%
%Summing over 3 families, the result is
%%
%\be
%\beta_{g_1} = \frac{41g_1^3} {6(4\pi)^2}.
%\ee

\paragraph{Yukawa coupling} 

The running of the top Yukawa follows from the counterterm $\delta_{y} =
\delta_{y_{h\bar \psi \psi}}-1/2(\delta_\phi +\delta_{t_L}
+\delta_{t_R})$. Explicit expressions can e.g. be found in \cite{zhou,
tubitak}.
For the top quark the SM counterterms are:
\begin{align}
\delta_{t_L} &=-\frac{1}{8\pi^2\eps} \(
\frac{\fac}{4} y_t^2 + \frac14 y_t^2 + \frac12 y_b^2\),
\nn \\
\delta_{t_R} &=-\frac{1}{8\pi^2\eps} \(
\frac{\fac}{4} y_t^2 + \frac14 y_t^2 + \frac12 y_t^2\).
\end{align}
The $y^2$ contributions stem from loops with $h$, $\chi$ and
$\phi^+ = \chi_1-i\chi_2$ respectively.  For $t_L$ the $\phi^+$ loop can
only be made with a bottom quark in the loop, which gives a
contribution to $y_b^2$ --- which we neglect in the following.  In our U(1) toy model we only had the first two
contributions from the $h$ and $\chi$-loop; indeed this matches the
counterterm we found before. In the mid and large field regime the Higgs loop is
suppressed, but not the GB loop.  Likewise, we expect the
$\phi^+$-loop to contribute in the large field regime, as these are
GBs, with the same structure of interactions as $\chi$.

The Higgs counterterm is
\be
\delta_\phi =-\frac{1}{8\pi^2\eps} 
\[ n_c y_t^2 -3 \fac\sum_a g_a^2\]
=-\frac{1}{8\pi^2\eps} 
\[ 3 y_t^2 -\frac34 \fac ( 3 g_2^2 +g_1^2)\],
 \ee
 which generalizes our previous results.  Here $n_c$ is the number of
 colors.  The gauge contribution is suppressed in the large field
 regime.

The vertex correction is
\be
\delta_{y_{h\bar \psi \psi}}= -\frac{1}{8\pi^2\eps}  \[-\frac12y_t^2 (\fac-1) +3 Y_{tL} Y_{tR}
g_1^2+ 3 C_2(R_t)g_3^2\],
\label{delta_PhiQ}
\ee
with $C_2(R_t) =4/3$ for the fundamental in SU(3), and $Y_{tL}=1/6,\, Y_{tR}=2/3$.
For the U(1) model we found that the $y^2$-correction from the $h$ and
$\chi$-loop cancels in the small field regime. However, in the
large field regime the Higgs contribution is negligible, and there is
a net contribution from the GB $\chi$. The $\phi^+$ loop gives a
contribution $\propto y_b^2$ and can be neglected.
The gauge terms stem from top-top-gauge  loops, and since the
fermion-gauge couplings are standard in the large field regime, these are
unaffected.

This gives for the betafunction
\be
\beta_{y_t} = \frac1{(4\pi)^2} \[\frac32 (2+\fac) y_t^3
-\( \frac{8+9\fac }{12} g_1^2  +\frac{9\fac}{4} g_2^2 +8g_3^2
\)y_t\].
\ee
In the small field regime $\fac =1$ and we get the standard result.

\paragraph{Non-minimal coupling}
Further, we found in the large and mid field regime that $\delta_\phi
=-\delta_\xi$. This gives the betafunction for the non-minimal coupling
\be
\beta_\xi \big|_{\rm mid, large}
= \frac{1}{(4\pi)^2} \[ 6 y^2 -\frac32 \fac  ( 3 g_2^2 +g_1^2)\]\xi.
\ee

%%%%%%%%%%%%%%%%%%%%%%%%%%%%%%%%%%%%%%%%%%%%%%%%%%
\subsubsection{End results}

\begin{align}
(4\pi)^2 \beta_\lambda 
&= 24 \lambda^2\fac  + A
+(4\pi)^2 \cdot 4\lambda \gamma_\phi \nn \\
(4\pi)^2\gamma_\phi 
&= -\frac{\fac}{4} (3g_1^2+9 g_2^2) + 3y_t^2
\nn \\
(4\pi)^2\beta_{g_3} &= -7 g_3^3,\nn\\
(4\pi)^2\beta_{g_2} &= - \frac{(20 - \fac)}{6}g_2^3, \nn\\
(4\pi)^2\beta_{g_1} &=\frac{(40 + \fac )}{6}g_1^3 \nn \\
(4\pi)^2\beta_{y_t} &= \frac32 \fac y_t^3
-\( \frac{2}{3} g_1^2  +8g_3^2
\)y_t+(4\pi)^2 \cdot\gamma_\phi  y_t\nn\\
(4\pi)^2\beta_\xi\big|_{\rm mid, large} &= (4\pi)^2 \cdot2 \gamma_\phi \xi
\label{final}
\end{align}
with $A= (3/8)(2g_2^4 + (g_2^2+g_1^2)^2) -6y_t^4$.

%%%%%%%%%%%%%%%%%%%%%%%%%%%%%%%%%%%%%%%%%%%%%%%%%%%%%%%%%%%%%%
%%%%%%%%%%%%%%%%%%%%%%%%%%%%%%%%%%%%%%%%%%%%%%%%%%%%%%%%%%%%%%
%%%%%%%%%%%%%%%%%%%%%%%%%%%%%%%%%%%%%%%%%%%%%%%%%%%%%%%%%%%%%%

\section{Discussion}
\label{s:concl}

\subsection{Comparison with literature.}  
%Only expressions for large field regime not
%for mid field.
%
%Our result in the large field regime
%%
%%
%\begin{align}
%(4\pi)^2 \beta_\lambda 
%&= 24 \lambda^2\fac  + A
%- 4\lambda \gamma_\phi \nn \\
%(4\pi)^2\gamma_\phi 
%&= \frac{\fac}{4} (3g_1^2+9 g_2^2) - 3y^2
%\nn \\
%(4\pi)^2\beta_{g_3} &= -7 g_3^3,\nn\\
%(4\pi)^2\beta_{g_2} &= - \frac{(20 - \fac)}{6}g_2^3, \nn\\
%(4\pi)^2\beta_{g_1} &=\frac{(40 + \fac )}{6}g_1^3 \nn \\
%(4\pi)^2\beta_{y_t} &= \[\frac32 \fac y_t^3
%-\( \frac{2}{3} g_1^2  +8g_3^2
%\)y_t\] -\gamma_\phi  y\nn\\
%(4\pi)^2\beta_\xi &= -2 \gamma_\phi \xi
%\end{align}
%%
%with $A= (3/8)(2g_2^4 + (g_2^2+g_1^2)^2) -6y^4$.  The QCD
%beta-function unaffected by the non-minimal coupling.

In recent years several groups have presented renormalization group equations for SM Higgs inflation. The disagreement between these results has been a major motivation to write this paper which follows, in our opinion, the most systematic approach so far. In this section we compare our findings to some encountered in the recent literature.

Let us first quickly compare this work to our own previous work \cite{damien}. There we have studied the renormalization of just a (complex) non-minimally coupled scalar, leaving the inclusion of fermions and gauge fields and the generalization to full SM Higgs inflation to this work. Our findings here generalize those in \cite{damien}. Note in particular that in that work, we concluded that we could not say anything about the RG flow in the mid-field regime, as the corrections were an order of $\delta$ smaller than the counterterms. However, the fermionic and some of the gauge corrections that we have found now are of the same order as the counterterms. That is why we can now present expressions for the running couplings in the mid field regime (barring threshold corrections) without getting in contradiction with our previous work.

Now for the comparison to other authors. In general, it seems that all other approaches follow some predefined treatment for Higgs and Goldstone bosons. In some cases only the Higgs contributions are kept, in other cases only the GBs, in yet other cases only loop contributions are excluded, etcetera. Our result does not respect any of these guidelines. For example, we find that the GB contribution to the effective potential is suppressed, while the GB loops do contribute to the Yukawa corrections. To see exactly which field contributes to which loop correction all correction diagrams need to be properly computed. (Although the differences are never very dramatic.)

We first compare to reference \cite{bezrukov3}, which states that in large field the action is just the
action of the chiral SM with $v = \mpl/\sqrt{\xi}$, and the
Higgs (but not the GBs) decouples.  The RGEs quoted for the large field
regime are
\begin{align}
(4\pi)^2 \beta_\lambda 
&=  A
+ (4\pi)^2 \cdot4\lambda \gamma_\phi \nn \\
(4\pi)^2\gamma_\phi 
&= -\frac{1}{4} (3g_1^2+6 g_2^2) + 3y_t^2
\nn \\
(4\pi)^2\beta_{g_2} &= - \frac{(20 - 1/2)}{6}g_2^3, \nn\\
(4\pi)^2\beta_{g_1} &=\frac{(40 + 1/2)}{6}g_1^3 \nn \\
(4\pi)^2\beta_{y_t} &= 
-\( \frac{2}{3} g_1^2  +8g_3^2
\)y_t +(4\pi)^2 \cdot\gamma_\phi  y_t\nn\\
(4\pi)^2\beta_\xi &= (4\pi)^2 \cdot 2 \gamma_\phi \xi.
\end{align}
Here a gauge contribution in $\gamma_\phi$ has been included, which explains the
difference in $\beta_\lambda$, $\beta_\xi$ and $\beta_{y_t}$ with
our work.  In the betafunctions for the gauge coupling only the real Higgs
field is excluded, whereas we also exclude the GB. 
% As for the Yukawa betafunctions, both
%the Higgs and GB contribution are excluded from $\beta_{y_t}$.

The chiral SM is non-renormalizable, and new operators have to be
included. At one-loop level there is a correction to the Z-boson mass,
which depends on the running coefficients of two of these new
operators. There is no such thing in our set-up, which is renormalizable in the
EFT sense.

We furthermore note that in this work the running of $\xi$ is computed with an approach very different from ours, via the running of the SM vev $v$. However, apart from our disagreement over the gauge contribution to $\gamma_\phi$, we find the same answer.

Reference \cite{wilczek} states that in the inflationary regime quantum loops
involving the Higgs field are heavily suppressed.  The proposed prescription (originally introduced in \cite{Salopek:1988qh}) is to
assign one factor of $s(\phi)$ for every off-shell Higgs that runs in
a quantum loop, with
\be
s(\phi) = \frac1{\Omega^2} \gamma_{\phi\phi}
= \frac{1+ \xi \phi_0^2}{1+(6\xi+1) \xi \phi_0^2}.
\ee 
Although the metric factor in $s$ above is for the real Higgs field,
judging from the RGEs presented the prescription has been applied to the
complex field (nothing is said explicitly about Higgs and GBs).
In large field, the quoted RGEs reduce to:
\begin{align}
(4\pi)^2 \beta_\lambda 
&= 
%24 \lambda^2s^2  + 
A
+ (4\pi)^2 \cdot4\lambda \gamma_\phi \nn \\
(4\pi)^2\gamma_\phi 
&= -\frac{1}{4} (3g_1^2+9 g_2^2) + 3y_t^2
\nn \\
(4\pi)^2\beta_{g_2} &= - \frac{20}{6}g_2^3, \nn\\
(4\pi)^2\beta_{g_1} &=\frac{40 }{6}g_1^3 \nn \\
(4\pi)^2\beta_{y_t} &= 
%\[
%(-\frac92 s +
-3
%)
y_t^3
-\( \frac{2}{3} g_1^2  +8g_3^2
\)y_t
%\] 
+
(4\pi)^2 \cdot\gamma_\phi  y_t\nn\\
(4\pi)^2\beta_\xi &=
%(2\lambda s+
 2 ((4\pi)^2 \cdot\gamma_\phi + 6\lambda)
 %)
 ( \xi+1/6).
\end{align}
There is a gauge contribution to $\gamma_\phi$, which explains the
difference in $\beta_\lambda$, $\beta_\xi$ and partly $\beta_{y_t}$ with
our result. 
 %For $\beta_y$ the $y^2$-term in
%$\gamma_\phi$ seems to be excluded as well. 
In $\beta_{g_i}$ the full Higgs doublet is
taken out in the large field regime, in agreement with our result.
$\beta_\xi$ is found by taking gravity as a classical background, following the pioneering work in \cite{Odintsov1,Odintsov2,Odintsov3}.  We think that this is not
a good approximation in the Jordan frame.
 %(they also use the above expression for
%the small field regime, where i don't think it's correct). 
%No explicit expression for $\beta_\xi $ in the large field regime is given,
%but from the statement that the $\lambda$-term is suppressed in the large field
%regime, we infer that it should be of the above form.  

This same s-factor formalism was followed in References \cite{rose, kyle}, with the modification that now only the loops of the real Higgs field are excluded, and not those of the GBs. However our final answers agree with neither of the results obtained there.

Reference \cite{barvinsky3} writes that Goldstone modes, in contrast to the
Higgs particle, do not have mixing with gravitons in the kinetic
term. Therefore, their contribution is not suppressed by the
$s$-factor.  We disagree with this. The GBs cannot be treated as
usual, as in polar coordinates $\rho^2 (\partial \theta)^2$ the radial
field is not the canonical one. In Cartesian coordinates, all
fields are equally coupled to the Ricci tensor via $\xi R \sum
(\phi^i)^2$. The quoted results are
\begin{align}
(4\pi)^2 \beta_\lambda 
&= 6 \lambda^2 + A
- (4\pi)^2 \cdot 4\lambda \gamma_\phi \nn \\
(4\pi)^2\gamma_\phi 
&= \frac{1}{4} (3g_1^2+9 g_2^2) - 3y_t^2
\nn \\
(4\pi)^2\beta_{g_2} &= - \frac{(20 - 1/2)}{6}g_2^3, \nn\\
(4\pi)^2\beta_{g_1} &=\frac{(40 + 1/2)}{6}g_1^3 \nn \\
(4\pi)^2\beta_{y_t} &= \[ y_t^3
-\( \frac{2}{3} g_1^2  +8g_3^2
\)y_t\] -(4\pi)^2 \cdot\gamma_\phi  y_t\nn\\
(4\pi)^2\beta_\xi &=  6\xi \lambda -(4\pi)^2 \cdot2\gamma_\phi \xi.
\end{align}
A gauge contribution to $\gamma_\phi$ is included, which partly explains
the difference in $\beta_\lambda$, $\beta_\xi$ and $\beta_{y_t}$ with
our results.  For $\beta_{y_t}$  the contribution of one GB
$y_t^2$-term has been excluded instead of 3GB $y_t^2$-terms. In $\beta_{g_i}$ only the GB is
taken out in the large field regime, in disagreement with our result.
%Here again $\beta_\xi$ is found by taking gravity as a classical background, which we think is not a good approximation in the Jordan frame.

% Not
%a good approx in the J-frame (they also use the above expression for
%the small field regime, where i don't think it's correct); in large
%field they only exclude GB, instead of full Higgs (also i think
%missing there are 3 GBs).
%

This should be the identical to the chiral model of \cite{bezrukov3}, as
the Higgs field is decoupled in the large field limit.  However the RGEs are still different.

Lastly, \cite{bezrukov3,wilczek} use two different normalization conditions,
one with a field independent cutoff in the Jordan frame, or with a
field dependent cutoff in the Jordan frame. However, two frames give identical physics. It is often quoted that a
field independent cutoff in the Jordan frame corresponds to a field
dependent cutoff in the Einstein frame, and vice versa.  However,
dimensionful quantities by themselves have no invariant meaning,
their values depend on the unit system.  If we express the cutoff in
Planck units (the Planck mass is frame dependent), a constant cutoff in the
one frame is equivalent to a constant cutoff in the other frame. The
conformal rescaling only rescales {\it all} length scales, which does
not change the physics. See also our discussion in \cite{damien}. On-shell equivalence between the frames has also been established in \cite{christian}.

The question about a field dependent or independent cutoff is a frame
invariant question when expressed in Planck units.  The choice of
cutoff has thus nothing to do with a choice of frame.  One can still
debate whether the results depend on the (field dependent) choice of
cutoff in the Einstein frame. A priori, this is not expected; the
cutoff is only introduced to regularize the divergent integrals, but
is at the end taken to infinity.  Requiring the counterterms to be
field independent, the different field dependent and independent
cutoffs lead to different normalization conditions.  In practice one
can only relate physical measurements at different energy scales. The
translation between the observable and the coupling defined in the
normalization condition will be different in each case. The end result
is that when comparing physical observables at different scales, the
cutoff dependence drops out.

\subsection{Conclusions}

We have calculated the one-loop corrections to Higgs inflation in the
small, mid and large field regime.  We have done the calculations for
the Abelian Higgs model; the results can then rather straightforwardly
be generalized to full Standard Model Higgs inflation. We have found
that in all three regimes the model is renormalizable in the effective
field theory sense.  The RGEs for SM Higgs inflation we found are
given in \eref{final}. The results for the mid field regime are new.
The running of the non-minimal coupling can be derived in the mid and
large field regime, and follows from the consistency of the radiative
corrections to the potential and to the two-point functions. In the
small field regime all dependence on the non-minimal coupling drops
out of the equations at leading order in the small field expansion,
and nothing can be said about its running.

The computation of the radiative corrections was done in the Einstein frame, in the Landau gauge, using a covariant formalism for the multi-field system. The one-loop corrections to the propagators are sufficient to determine the full set of counterterms, and thus the betafunctions. As extra checks, we have calculated many higher n-point functions as well. Especially the results for four-point scattering of the Goldstone bosons are impressive in this regard: in the large (and mid) field regime both the leading and subleading divergencies exactly cancel, yielding a consistent counterterm.

However, we have stumbled on some potential problematic outcomes as well.  First of all, to cancel all divergencies, new non-renormalizable counterterms need to be added, see (\ref{CU1},~\ref{CU2}).  However, the cutoff implied by these new counterterms always exceeds the unitarity cutoff (\ref{unitarity}). Therefore they do not put further constraints on the validity of the EFTs in the various regimes.  Second, in (\ref{transversal}) we have seen that the gauge boson propagator picks up a transverse part, that should be absent by the Ward identities. This term vanishes in the large $\xi$ limit. Moreover, it is gauge dependent, and we believe it should vanish in a full calculation. Thirdly, the one-loop gauge-fermion vertex gives a gauge symmetry breaking result as well. We can trace it back to the explicit symmetry breaking in our approximation of the non-minimal kinetic terms. In a full calculation the symmetry should be restored, yielding a result consistent with the gauge propagator corrections that we have found, but we leave this for further work.

In conclusion, we have computed the full set of RGE equations for Standard Model Higgs inflation.  The Higgs-fermion part has withstood an impressive set of consistency checks. When including the gauge symmetry our result obtained from propagator corrections has failed one consistency test, which we think can be ascribed to the intrinsic limitations in our approach (neglecting higher order kinetic terms by evaluating the field metric on the background). It would be an interesting but equally challenging task to develop a framework that can get around these limitations.

% However, this should not affect the simple diagrams making up the gauge propagator corrections and, therefore, the form of the RGEs that we have found in this work.
%
%
%
% there remain some questions which deserve further study.
%
%In conclusion, we have computed the full set of RGE equations for Standard Model Higgs inflation.  The Higgs-fermion part withstood an impressive set of consistency checks, and we believe these results can be fully trusted. When including the gauge symmetry there remain some questions which deserve further study.
%

\section*{Acknowledgments}

DG is funded by a Herchel Smith fellowship.
SM is supported by the Fondecyt 2015 Postdoctoral Grant 3150126 and by the ``Anillo'' project ACT1122, funded by the ``Programa de Investigaci\'on Asociativa."
MP is funded by the Netherlands Foundation for
Fundamental Research of Matter (FOM) and the Netherlands
Organisation for Scientific Research (NWO). 
We thank Mikhail Shaposhnikov and Sergey Sibiryakov for illuminating discussions.

\appendix
\section{Couplings} 
\label{s:vertices}

We list the couplings for the Abelian U(1) model.  The explicit values
in the small, mid, and large field regime are found expanding in
$\delta_i$:
\begin{align}
& \delta_s = \xi \phi_0, & {\rm small} \nn \\
&\xi \to \delta_m^{-2}\xi,\;\phi_0 \to
\delta_m^{3/2} \phi_0, & {\rm mid} \nn \\
&  \delta_l = 1/(\xi \phi_0^2) . & {\rm large}
\end{align}
Note that contrary to the small and large field regime, the $\delta_m$
parameter is just a rescaling parameter. In the small field we express
the results in $\phi_0$ rather than $\delta_s$, as this form is more
familiar.  Below we give the leading expression for the metric and for the
relevant couplings; the three values between the braces correspond to
the small, mid and large field regime.

The metric is:
\begin{align}
\gamma_{hh} & %=\frac{\phi_0^2 \xi  (6 \xi +1)+1}{\Omega_0^4}
=\left\{1,~\frac{6 \phi_0^2 \xi ^2}{\delta _m}, ~(6 \xi +1) \delta
   _l\right\}
\nn \\
\gamma_{\chi\chi} & %=\frac{1}{\Omega_0^2}
=\left\{1,~1,~\delta _l\right\}.
\end{align}

The Higgs and GB self-interactions are
\begin{align}
\lambda_{2h} &= \frac{1}{2!} V_{;\phi\phi} \Big|_{\rm bg} 
=\lambda \left\{ \frac32 \phi_0^2, ~ \phi_0^2 \delta_m^3, ~ -\frac1\xi
               \delta_l^2 \right\}
\nn\\
\lambda_{2\chi} &=
\frac{1}{2!} V_{;\theta\theta} \Big|_{\rm bg} 
=\lambda \left\{ \frac12 \phi_0^2, ~ \frac{1}{12\xi^2} \delta_m^4, ~ \frac{
               \delta_l^3}{2\xi(1+6\xi)} \right\}
\nn\\
\lambda_{3h} &= \frac{1}{3!} V_{;\phi\phi\phi} \Big|_{\rm bg} =\lambda
\left\{
 \phi_0,~ \(\frac1{18\phi_0 \xi^2} -2\phi_0^3 \xi\) \delta_m^{5/2},
~ \frac{2\delta_l^{5/2}}{3\sqrt{\xi}} \right \}  
\nn\\
\lambda_{h2\chi} &= \frac{1}{3!} \left(V_{;\phi\theta\theta} +
                     V_{;\theta \phi \theta} + V_{;\theta \theta \phi}
                     \right)  \Big|_{\rm bg} 
=\lambda\left\{\phi_0, ~ \frac{\delta_m^{5/2}}{18 \phi_0 \xi^2},~
-\frac{4 \delta_l^{7/2}}{3\sqrt{\xi}(1+6\xi)} \right\}  
\nn\\
\lambda_{4h} &= \frac{1}{4!}V_{;\phi\phi\phi\phi} \Big|_{\rm bg} = 
\lambda\left\{ \frac14,~ -\frac{\delta_m}{18\phi_0^2 \xi^2}, ~
~- \frac{\delta_l^3}{3} \right\}
 \nn\\
\lambda_{2h2\chi} &= \frac{1}{4!} \left( V_{;\phi\phi\theta\theta} +
                    {\rm 5 perms}  \right)   \Big|_{\rm bg} 
=\lambda \left\{ \frac12, ~ -\frac{\delta_m}{18\phi_0^2\xi^2},~
                    \frac{11 \delta_l^4}{6(1+6\xi)} \right\}
\nn\\
\lambda_{4\chi} &= \frac{1}{4!}V_{;\theta\theta\theta\theta}
                  \Big|_{\rm bg} 
=\lambda \left\{\frac14,~ \frac{\delta_m^2}{432 \phi_0^4 \xi^4},~
                  -\frac{\delta_l^5}{3(1+6\xi)^2}
\right\}
\nn\\
\lambda_{5h} &= {\rm etc.}
\end{align}

The Yukawa interactions are
\begin{align}
m_\psi & =F^\phi  \Big|_{\rm bg} 
=\frac{y}{\sqrt{2}} \left\{\phi_0 ,~\phi_0 \delta
   _m^{3/2},~\frac{1}{\sqrt{\xi }}\right\}
\nn \\
y_h & =F^\phi_{;\phi}  \Big|_{\rm bg} 
=\frac{y}{\sqrt{2}} \left\{1,~1,~\delta_l^{3/2}\right\}
\nn\\
y_{2h} &= \frac1{2!}F^\phi_{;\phi\phi}  \Big|_{\rm bg}  
=\frac{y}{\sqrt{2}} \left\{-(3 \xi^2 +\xi) \phi_0,~ -\frac{1}{2\phi_0
         \delta _m^{3/2}}, -\sqrt{\xi } \delta
   _l^2 \right\}
\nn \\
y_{3h} & =\frac{1}{3!}F^\phi_{;\phi\phi\phi}  \Big|_{\rm bg} = 
\frac{y}{\sqrt{2}} \left\{-\frac13 (3 \xi^2 +\xi),~-\frac{1}{2 
   \phi_0^2 \delta _m^{3}},~\frac23 \xi \delta_l^{5/2} \right\}
\nn \\
y_{4h} &= {\rm etc.} \nn \\
y_\chi &=F^\theta_{;\theta}  \Big|_{\rm bg}  =
\frac{y}{\sqrt{2}}\left\{1,~1,~ \sqrt{\delta_l} \right\}
\nn \\
y_{2\chi} &=\frac1{2!} F^\phi_{;\theta\theta}  \Big|_{\rm bg}  
=\frac{y}{\sqrt{2}}\left\{-(3 \xi^2 +\xi) \phi_0 ,~-\frac{1}{2 \phi_0
            \delta _m^{3/2}},~-\frac12\sqrt{\xi } \delta _l \right\}
\nn \\
y_{3\chi} &=\frac{1}{3!}F^\theta_{;\theta\theta\theta}  \Big|_{\rm bg}  =
 \frac{y}{\sqrt{2}} \left\{-\frac13 (3 \xi^2 +\xi),~-\frac{1}{6 
   \phi_0^2 \delta _m^{3}},~-\frac16 \xi \delta_l^{3/2} \right\}
\nn \\
y_{4\chi} &=\frac1{4!}F^\phi_{;\theta\theta\theta\theta}  \Big|_{\rm bg}  =
            \frac{y}{\sqrt{2}} \left\{\(\frac{\xi^2}{3} +3 \xi^3 +\frac{15\xi^4}{2}\)\phi_0,~\frac{1}{24
   \phi_0^3 \delta _m^{9/2}},~\frac1{24} \xi^{3/2} \delta_l^{2} \right\}
\end{align}
with 
\be
F^\phi = \frac{y}{\sqrt{2}} \frac{\phi_0 + \vp}{\Omega}, \qquad
F^\theta = \frac{y}{\sqrt{2}} \frac{\theta}{\Omega},
\ee
as follows from \eref{VF}.
 
Finally, the gauge interactions are
\begin{align}
m_A^2 &  =g^2 G \Big|_{\rm bg}  
= g^2\left\{\ \phi_0^2, ~\phi_0^2 \delta_m^3, ~ \frac1\xi \right\}
\nn \\
g_{2Ah} & =g^2 G_{;\phi} \Big|_{\rm bg}  =
g^2\left\{\ \phi_0, ~\phi_0 \delta_m^{3/2}, ~ \frac{\delta_l^{3/2}}{\sqrt{\xi}} \right\}
\nn \\
g_{2A2h} & =\frac1{2!} g^2 G_{;\phi\phi} \Big|_{\rm bg}  =
g^2\left\{\ \frac12, ~\(\frac1{12\phi_0^2 \xi^2} -\phi_0^2\xi\)\delta_m,
 ~ -\delta_l^{2} \right\}
\nn\\
g_{2A2\chi} & =\frac1{2!} g^2 G_{;\theta\theta} \Big|_{\rm bg}  =
g^2\left\{\ \frac12, ~\frac{\delta_m}{12\phi_0^2 \xi^2},
 ~ \frac{\delta_l^{3}}{2(1+6\xi) }\right\}
\nn\\
g_{A \chi \partial  h } &=  g G^\phi_{;\theta} \Big|_{\rm bg}  
=g\left\{1,~1,~\delta _l \right\}
\nn \\
g_{Ah \partial \chi } &=  g G^\theta_{;\phi} \Big|_{\rm bg}  
=g\left\{1,~1,~-\delta _l\right\}
\nn\\
g_{A \bar \psi_{L,R} \psi_{L,R}} & = g q_{L,R} 
\end{align}
with $G,G^\phi,G^\theta$ defined in \eref{VF}. To get the two derivative couplings, we have simply set $\partial \phi=\partial h$ and $\partial \theta=\partial\chi$. For higher derivative couplings (to be precise: with still one derivative but more fields, such as $g_{A 2 h  \partial \chi }$) we need to go beyond the first terms in these expansions, by using \eref{Q_expl}.

%%%%%%%%%%%%%%%%%%%%%%%%%%%%%%%%%%%%%%%%%%%%%%%%%%%%%%%%%%%%%%%%
%%%%%%%%%%%%%%%%%%%%%%%%%%%%%%%%%%%%%%%%%%%%%%%%%%%%%%%%%%%%%%%%
%%%%%%%%%%%%%%%%%%%%%%%%%%%%%%%%%%%%%%%%%%%%%%%%%%%%%%%%%%%%%%%%

%\end{document}

\end{document}